\documentclass[aps,prb,twocolumn,showpacs,amsmath,amssymb,superscriptaddress,floatfix]{revtex4-2}
\usepackage[english]{babel}
\usepackage{blindtext}
\usepackage[utf8]{inputenc}
\usepackage{graphicx}
\usepackage{bm}
\usepackage{latexsym}
\usepackage{amsfonts}
\usepackage{bbold}
\usepackage{appendix}
\usepackage[colorlinks = true,
            linkcolor = blue,
            urlcolor  = blue,
            citecolor = blue,
            anchorcolor = blue]{hyperref}
\usepackage{ulem}
\usepackage{epstopdf}
\begin{document}

\author{Xianxin Wu}
\email{xxwu@itp.ac.cn}
\affiliation{CAS Key Laboratory of Theoretical Physics, Institute of Theoretical Physics, Chinese Academy of Sciences, Beijing 100190, China}
\author{Debmalya Chakraborty}
\affiliation{Department of Physics and Astronomy, Uppsala University, Box 516, S-751 20 Uppsala, Sweden}
\author{Andreas P. Schnyder}
\affiliation{Max Planck Institute for Solid State Research, Heisenbergstrasse 1, D-70569 Stuttgart, Germany}
\author{Andr\'es Greco}
\affiliation{Facultad de Ciencias Exactas, Ingenier\'{\i}a y Agrimensura and Instituto de F\'{\i}sica Rosario (UNR-CONICET), Av. Pellegrini 250, 2000 Rosario, Argentina}

\title{Crossover between electron-electron and  \\
electron-phonon mediated  pairing on the Kagome lattice}%
\date{\today}
\begin{abstract}
We study electron-electron and electron-phonon mediated pairing in the Holstein extended Hubbard model on the Kagome lattice near the van Hove fillings, and investigate their combined effects on electron pairing states. We find that their combination can promote exotic pairings in a crossover region, where the filling is close to a van Hove singularity. In particular, at the $p$-type van Hove filling the $E_{1u}$ ($p$-wave) and  $B_{2u}$ ($f_{y^3-3yx^2}$-wave) pairings become leading, and at the $m$-type van Hove filling the $E_{1u}$ and $A_{2g}$ ($i$-wave) pairings get promoted. Moreover, we show that the electron-phonon interaction acquires a significant momentum dependence, due to the sublattice texture of the Fermi surfaces, which can promote non $s$-wave pairing. We present a detailed analysis of these pairing propensities and discuss implications for the vanadium-based kagome superconductors AV$_3$Sb$_5$.
\end{abstract}
\maketitle

One of the central questions in the field of correlated superconductors
is the nature of the pairing mechanism,
which can be either electron-phonon driven or entirely electronic.
The answer to this question, which has been hotly debated in numerous classes of superconductors~\cite{Scalapino2012,chubukov_review_Fe_based}, directly determines the properties and topologies of the superconducting state:
While electron-phonon mediated pairing tends to produce trivial $s$-wave states~\cite{ALLEN19831,carbotte_review,brydon_sarma_phonon_pairing},
purely electronic pairing leads to topological states with higher angular momenta and possibly
nontrivial spin configurations~\cite{Schnyder_2015,Sato_2017}. The latter types of pairings are highly sought after, since
they have applications in quantum information and sensor technology~\cite{RevModPhys.80.1083}.

The recent discovery of superconductivity in Kagome materials, including  vanadium based AV$_3$Sb$_5$(A=K,Rb,Cs)~\cite{AV3Sb5_Ortiz_first_paper,PhysRevLett.125.247002,AV3Sb5_nature_review,jiangping_hu_review}, ruthenium based RERu$_3$Si$_2$ (RE=rare earth elements)~\cite{BARZ19801489,PhysRevMaterials.5.034803,ChinPhysLett.39.087401}, Ti-based materials~\cite{YangHT2022}
 and Ta$_2$V$_{3.1}$Si$_{0.9}$~\cite{LiuHX2023}, has brought the question of the pairing mechanism to the table again, but now with several new twists,
related to the sublattice texture and van Hove singularities of the Kagome lattice~\cite{SYu2012,PhysRevB.86.121105,WWang2013,PhysRevLett.110.126405,wu_PRL_21}. Especially, the vanadium based kagome material AV$_3$Sb$_5$ has attracted tremendous attention, owing to its possible unconventional superconducting and charge density wave (CDW) orders. With multiple van Hove (VH) singularities in the vicinity of the Fermi level~\cite{HChen2021,HZhao2021,MKang2022,YHu2022}, superconductivity emerges in AV$_3$Sb$_5$ inside a CDW order with a transition temperature of $T_c \simeq$ 0.9-2.5 K at ambient conditions~\cite{PhysRevLett.125.247002,PhysRevMaterials.5.034801,QiangweiYin.37403}. As VH singularities (VHSs) carry large density of states, correlation effects are expected to be crucial in these kagome superconductors. So far, the conflicting experimental evidences about superconducting gaps renders the pairing mechanism elusive~\cite{2020arXiv201205898W,2021arXiv210208356Z,PhysRevLett.126.247001,PhysRevB.103.224513,Chen_2021,PhysRevLett.127.237001,PhysRevMaterials.6.L041801,2021arXiv211012651L,HChen2021,PhysRevLett.127.187004,WDuan2021,Mu_2021,2022arXiv220207713G,ZhengLX2022}. A significant residual thermal conductivity~\cite{2021arXiv210208356Z}, superconducting domes with external pressure~\cite{PhysRevLett.126.247001,PhysRevB.103.224513,Chen_2021} and charge doping~\cite{PhysRevLett.127.237001,PhysRevMaterials.6.L041801,2021arXiv211012651L}, and V-shaped gaps in scanning tunneling microscopy (STM) measurements~\cite{HChen2021,PhysRevLett.127.187004} suggest unconventional pairing. In contrast, penetration depth and nuclear magnetic resonance (NMR) measurements suggest an electron-phonon-coupling-driven $s$-wave pairing~\cite{WDuan2021,Mu_2021}. Moreover, $\mu$SR measurements~\cite{2022arXiv220207713G} reveal a transition from a nodal to nodeless gap with increasing pressure and suggest that the nodeless pairing breaks time-reversal symmetry, when the CDW order is supressed by pressure. Recent angle-resolved photoemission spectroscopy measurements identified clear kinks in both Sb $p$-orbital and V $d$-orbital bands from which an intermediate electron-phonon coupling (EPC) strength was determined~\cite{2022arXiv220702407Z}. Thus, both electronic interactions (EI) and EPC are believed to play crucial roles in promoting the exotic orders of AV$_3$Sb$_5$~\cite{FENG20211384,PhysRevLett.127.217601,PhysRevB.104.045122,PhysRevB.104.035142,PhysRevLett.127.046401,Tazai2022,Wen2022,HLuo2022,2022arXiv220106477H,2022arXiv220201902K}. Hence, in order to properly address the mechanism of superconductivity in AV$_3$Sb$_5$, one must consider electron-phonon and electron-electron interactions on equal footing.

Motivated by this, we study in this Letter the pairing states on the kagome lattice at VH fillings by including both electron-phonon coupling and electronic interactions within the random phase approximation (RPA). We find that the effective pairing interaction from an isotropic electron-phonon pairing obtains a substantial momentum dependence due to the VH sublattice texture and can promote non s-wave pairing.
Moreover, the combined electron-phonon and electron-electron pairing mechanisms lead to electron pairings in the $p$-wave ($E_{1u}$) and $f_{y^3-3yx^2}$-wave ($B_{2u}$) channels near the p-type VH filling and in the $E_{1u}$ and $i$-wave ($A_{2g}$) channels near the m-type VH filling. The mechanism is analyzed and implications for pairing in AV$_3$Sb$_5$ are discussed.

\begin{figure}[t!]
\begin{center}
\includegraphics[width=0.5\textwidth]{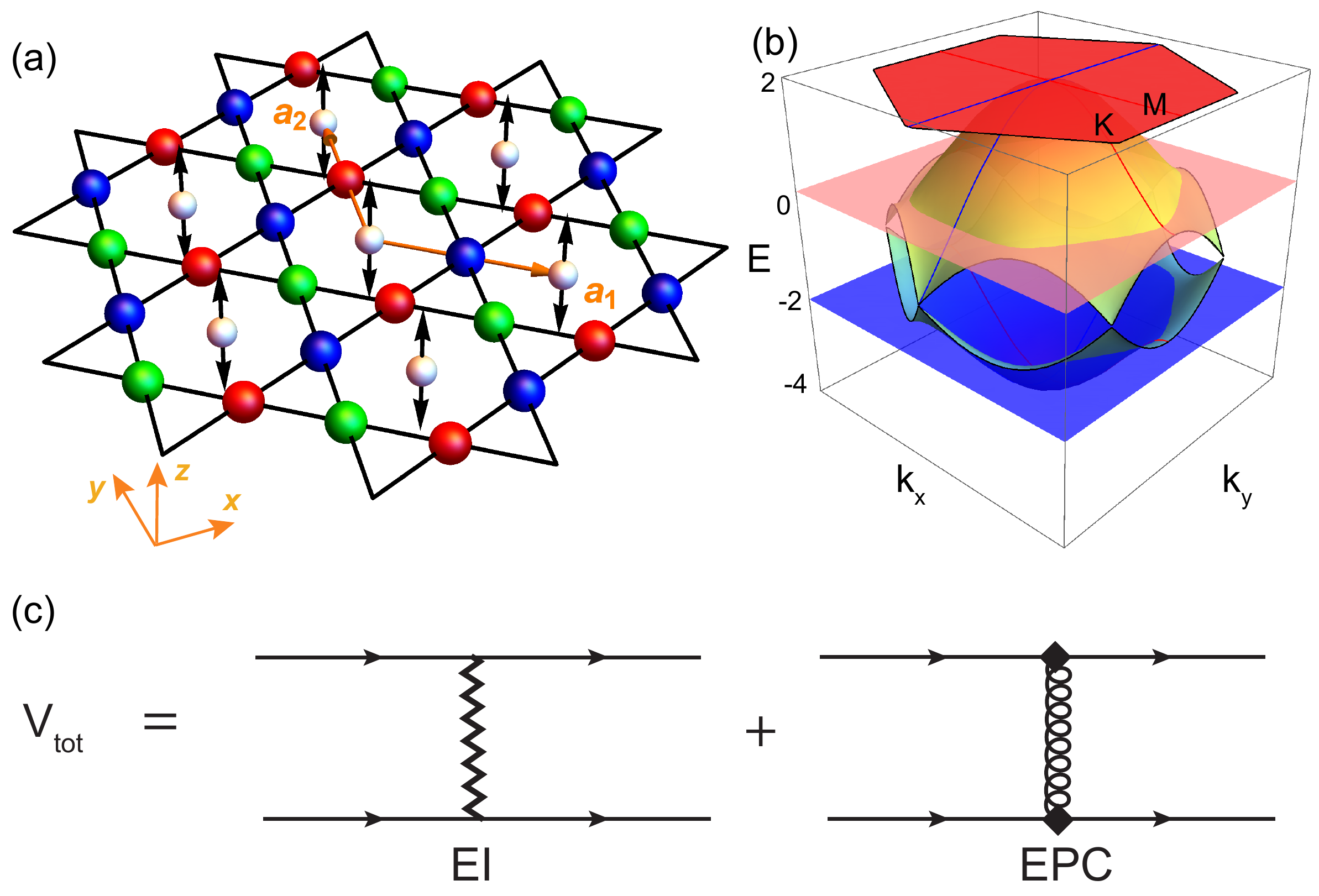}
\caption{
(a) Kagome lattice structure with the three sublattices A, B and C indicated by the red, green and blue spheres, respectively. $\bm{a}_1$ and $\bm{a}_2$ denote the unit vectors. The white spheres located at the center of hexagons with arrows represent the vibrations of the out-of-plane Holstein phonon mode.
(b) Band structure in the Brillouin zone with the pink and blue planes denoting fillings of p-type ($\mu=0.08$) and m-type ($\mu=-2.08$) near the van Hove singularities. The red and blue lines represents band dispersion along $\Gamma-M$ and $\Gamma-K$, respectively.
(c) Total effective interaction from the electronic interaction and electron-phonon coupling.
}
\label{fig_figure1}
\end{center}
\end{figure}

%
\textit{Model and formalism.--}
To explore the superconducting pairing on the kagome lattice (Fig.~\ref{fig_figure1}(a)), we consider a tight-binding model $\mathcal{H}_{0}$ with an extended ($U$-$V$) Hubbard interaction and a Holstein-type electron-phonon coupling. The  $U$-$V$ Hubbard Hamiltonian reads,
\begin{equation}
   \mathcal{H}_{\text{int}}=U\sum_{i,\alpha} n_{i\alpha\uparrow}n_{i\alpha\downarrow} + V\sum_{\substack{\langle ij \rangle,\alpha,\beta \\ \alpha \neq \beta}} n_{i\alpha} n_{j\beta},
   \label{eq:int}
\end{equation}
where $n_{i\alpha}=n_{i\alpha\uparrow}+n_{i\alpha\downarrow}$ is the electron density operator, $\langle ij \rangle$ denotes the nearest neighbor sites, $U$ is the onsite Hubbard repulsion and $V$ is the nearest neighbor Coulomb repulsion. For the phonons, we consider the Holstein model, which mimics the
out-of-plane vibration of the phonon subsystem at the center of the  hexagon in Fig.\ref{fig_figure1}(a). The study of superconductivity in interacting Holstein models is an old topic, and has resurfaced recently~\cite{Han20,Costa18,Marsiglio22}. The phonon and the electron-phonon part of the Holstein Hamiltonian is given by
\begin{eqnarray}
    \mathcal{H}_{\text{p}}+\mathcal{H}_{\text{ep}}&=&\sum_{\mathbf{q}}   \omega_{D}\left( a^{\dagger}_{\mathbf{q}}a_{\mathbf{q}}+ \frac{1}{2}\right)\nonumber \\
     &+&\frac{g_0}{\sqrt{N}}\sum_{\alpha\mathbf{k}\mathbf{q}\sigma} c_{\mathbf{k}+\mathbf{q}\alpha\sigma}^{\dagger}c_{\mathbf{k}\alpha\sigma}\left(a_{-\mathbf{q}}^{\dagger}+a_{\mathbf{q}}\right),
    \label{eq:pepmommain}
\end{eqnarray}
where $g_0$ is the bare electron-phonon coupling constant and $\omega_D$ is the phonon frequency, which can be associated to the Debye frequency for studying superconductivity.  $a^{\dagger}_{\mathbf{q}}$ ($a_{\mathbf{q}}$) is the phonon creation (annihilation) operator and $N$ is the total number of lattice sites. More details on the Hamiltonian are presented in the supplemental material (SM)~\cite{SM}. Defining the hopping parameter $t$ as the unit of energy, we set $t=1$ from now on. For the numerical calculations we choose the representative values $\omega_D=0.01$ and $g_0 =0.1$ (see SM~\cite{SM}).

\begin{figure*}[t!]
\begin{center}
\includegraphics[width=1.0\textwidth]{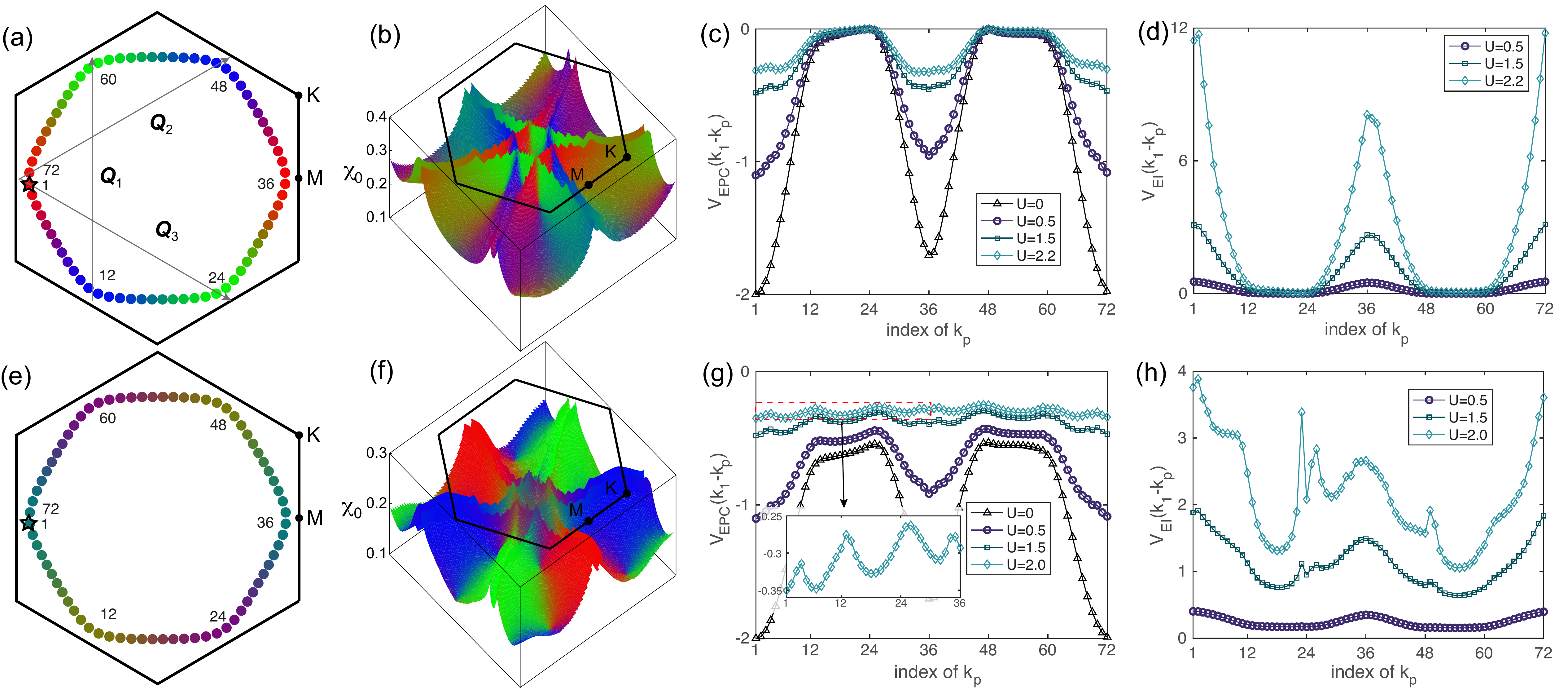}
\caption{
Sublattice-resolved Fermi surface and largest eigenvalues for the bare static susceptibility matrix $\chi_0(\bm{q})$ near the p-type VHS $\mu=0.08$ (a,b) and m-type VHS $\mu=-2.08$ (e,f), where the red, green blue color represent the weight of A, B and C sublattice, respectively.  $\bm{Q}_{1,2,3}$ are the nesting vectors. The effective Cooper pairing scattering interaction on the Fermi surface from the EPC and electronic interactions near p-type VHS (c,d), and m-type VHS $\mu=-2.08$ (g,h). The $\bm{k}_1$ is fixed near the M point denoted by a star in (a,e) and $\bm{k}_p$ runs on the Fermi surface. $V=0.3U$ in (c), (d), (g), and (h).
}
\label{fig_figure2}
\end{center}
\end{figure*}

 Diagonalizing the bare electron Hamiltonian we obtain three bands as shown in Fig.~\ref{fig_figure1}(b), one of which is flat and the other two form Dirac cones at the K points. At the M points there are two types of VHSs with energies $E=0$ and $E=-2$, featuring pure sublattice (p-type) or mixed sublattice (m-type) characters, respectively. The corresponding hexagonal Fermi surfaces near the two VH fillings are displayed in Fig.\ref{fig_figure2}(a) (p-type) and (e) (m-type), respectively. The red, green, and blue colors represent the weight of the A, B and C sublattices.

From the total Hamiltonian $\mathcal{H}=\mathcal{H}_0+\mathcal{H}_{\rm int}+\mathcal{H}_{\rm p}+\mathcal{H}_{\rm ep}$ (see SM), we have two contributions to the effective interactions for superconducting pairing: An electron-electron ($V_{\rm El}$) and an electron-phonon ($V_{\rm EPC}$) effective interaction, as shown diagrammatically in Fig.~\ref{fig_figure1}(c). In Fig.~\ref{fig_figure1}(c) the filled diamonds represent the electron-phonon coupling renormalized by correlations, a curly line indicates the bare phonon propagator and a darkened zig-zag line
represents the effective electron-electron interaction. We calculate the effective electronic interactions and renormalized electron-phonon coupling using the RPA, as detailed in SM. Near $T_c$, the gap function can be obtained by solving the linearized gap equation,
\begin{equation}
  -\int_{\rm FS} \frac{\sqrt{3}d {\bf k'}}{2(2 \pi)^2|v_{\bf k'}|}V^{\rm S/T}_{\eta}\left(\mathbf{k},\mathbf{k}^\prime\right) \Delta_{i}(\mathbf{k}^{\prime})=\lambda_i^ {\eta}\Delta_{i}(\mathbf{k}),
  \label{eq:gapeigenvalues}
\end{equation}
where $v_F(\textbf{k})$ is the Fermi velocity at the momentum ${\mathbf k}$ on the Fermi surface (FS). $\lambda^\eta_{i}$ denotes the pairing strength for the gap function $\Delta_{i}(\textbf{k})$ from the pairing interaction vertex $V_\eta$, in the triplet (T) and singlet (S) channels,  with $\eta=\text{EPC},\text{EI},\text{tot}$ (for details see SM).

\textit{Results and discussions.--}
At both VHSs, there are three nesting vectors $\bm{Q}_1=(0,2\pi)$ and $\bm{Q}_{2/3}=(\pm \sqrt{3}\pi,\pi)$, as shown in Fig.\ref{fig_figure2}(a). Around the p-type VHS, these nesting vector always connect distinct sublattice character of states around the three saddle points (Fig.\ref{fig_figure2}(a)). Around the m-type VHS, however, the Fermi surface segments near M points connected by $\bm{Q}_{1,2,3}$ always share a common sublattice. These features dominantly determine the intrinsic charge or spin fluctuations embedded in the susceptibility. Despite similar peak structures of the bare susceptibility near the $\Gamma$ and M points in the two cases, the corresponding distribution of sublattice weight in the momentum space are quite different, as shown in Fig.\ref{fig_figure2}(b) and \ref{fig_figure2}(f). Near the p-type VHS, the peaks around the $\Gamma$ point are mainly attributed to one sublattice while the peaks around the M point are attributed to a mixture of the other two sublattices, derived from the sublattice characters on the Fermi surface. In contrast, near the m-type, the distribution of sublattice weight of the susceptibility is the opposite. The peaks ascribed to one sublattice and mixed sublattices in the susceptibility can be enhanced by including onsite and nonlocal Coulomb interactions, respectively.
Due to the sublattice makeup on the Fermi surface, the effective interaction will be reduced in the band space. In the following, we will demonstrate the crucial effect of these salient sublattice features on the correlated phenomena.

We start with the case at a filling $\mu=0.08$ near the p-type VHS.
Here, only intra-sublattice pairing is allowed near three saddle points owing to the sublattice feature in Fig.\ref{fig_figure2}(a). We project the bare EPC interaction into the band space (see SM) and Fig.\ref{fig_figure2}(c) shows the pairing vertex $V_{\text{EPC}}(\bm{k}_1,\bm{k}_p)$, where $\bm{k}_1$ is fixed around the M point $(-\frac{2\pi}{\sqrt{3}},0)$ (marked by a star in Fig.\ref{fig_figure2}(a)) and $\bm{k}_p$ runs on the FS and is numbered from $1$ to $72$. Remarkably, we find that the interaction already exhibits a large anisotropy without including electronic interactions. It peaks at $\bm{k}_p=1,36$, where the eigenstates at both $\bm{k}_1$ and $\bm{k}_p$ are dominantly contributed by the A sublattice and rapidly falls to almost zeros at $\bm{k}_p=12,24,48,60$, where the eigenstates are mainly attributed to B and C sublattices.
Such sublattice dependence is intimately related to the sublattice characters at the saddle points and the density-type EPC, which can only generate an intra-sublattice Cooper pair scattering but no inter-sublattice one. This is in contrast to the case of the bare local coupling of the Holstein type which is usually expected to generate an isotropic effective attraction between electrons~\cite{Foussats_2006,Foussats_PRB_05}.

\begin{figure*}[t!]
\begin{center}
\includegraphics[width=0.8\textwidth]{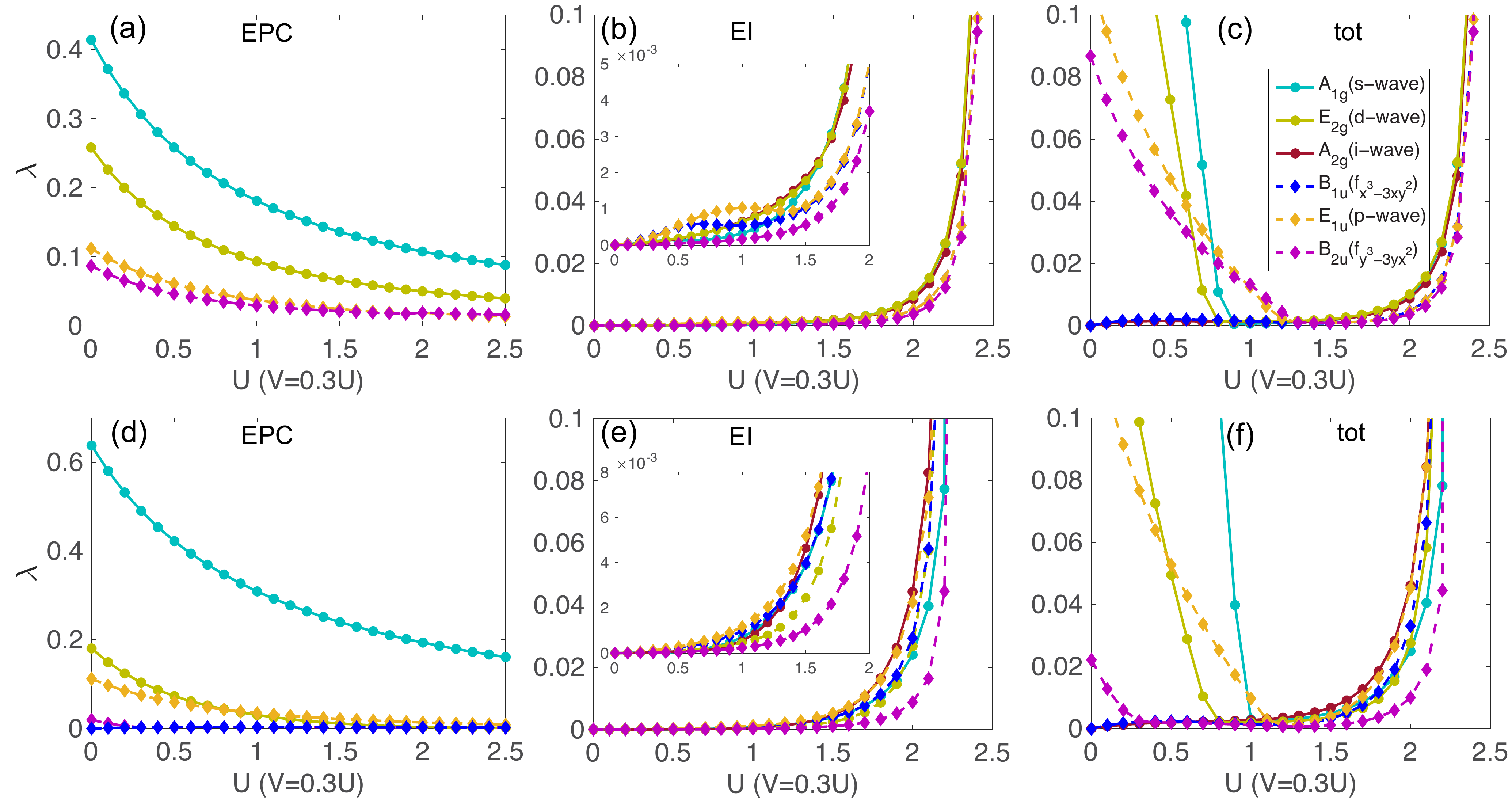}
\caption{
Pairing strength eigenvalues for the dominant instabilities as a function of U (V=0.3U) near the p-type (top panels) and m-type (bottom panels) and the contribution from EPC (a,d), electronic interactions (b,e) and their summation (c,f). Here, both bubble and ladder diagrams are included for EI. The results with only bubble diagrams are given in the SM.}
\label{fig_figure3}
\end{center}
\end{figure*}

Once electronic interactions are introduced, the EPC vertex will get renormalized and reduced, as shown in Fig.\ref{fig_figure2}(c), but the anisotropy remains. The effective interaction directly derived from the EI is displayed in Fig.\ref{fig_figure2}(d). Except for the repulsive nature, the general features are similar to the EPC. Here, the peak around $\bm{k}_p=1$ is dominantly contributed by the ladder diagrams as the spin fluctuations peak off the $\Gamma$ point and the sharp dip around $\bm{k}_p$ index $12,24,48,60$ is because spin or charge fluctuations with a momentum $\mathbf{Q}_{1,2,3}$ cannot mediate intersublattice Cooper scattering on the Fermi surface.

For the filling $\mu=-2.08$ near the m-type VHS, the effective interaction from the bare EPC is displayed in Fig.\ref{fig_figure2}(g). It is less anisotropic and especially it falls to a quarter of $V_{\text{EPC}}(\bm{k}_1,\bm{k}_1)$ for $\bm{k}_p=12,24,48,60$. This value is nonzero as the eigenstates at those $\bm{k}$ points share a same sublattice B or C with that at $\bm{k}_1$. With increasing electronic interactions, the EPC interaction becomes more isotropic and additional features appear, as shown in Fig.\ref{fig_figure2}(g). At $U =2$ shown in the inset, the interaction exhibits dips around $\bm{k}_p=12,24$, which is derived from the renormalization by the enhanced spin or charge fluctuations at $\bm{Q}_{2/3}$. The pairing vertex from the EI is displayed in Fig.\ref{fig_figure2}(h). It exhibits similar features but peaks at the dip positions of the EPC case as the corresponding fluctuations enhance electronic pairing vertex.

We further analyze the pairing propensity near two VH fillings. For $\mu =0.08$ EPC alone is expected to generate a dominant $s$-wave ($A_{1g}$) pairing due to its attractive nature, as shown in Fig.~\ref{fig_figure3}(a). Remarkably, even for the non-interacting case $U=V=0$ the bare EPC can already create competing pairings in the $d$-wave ($E_{2g}$), $p$-wave ($E_{1u}$), and $f_{y^3-3yx^2}$-wave ($B_{2u}$) channels, owing to the anisotropic pairing vertex. This is in sharp contrast to the vanishing non-$s$-wave pairing in the triangular or square lattice\cite{Foussats_2006,Foussats_PRB_05}. These subleading states are pairings between nearest-neighbor (NN) sites and the $B_{2u}$ pairing features a sign change under a 60$^{\circ}$rotation with line nodes along $\Gamma-M$ direction (see SM). With increasing $U$ and a fixed ratio $V/U=0.3$, the pairing vertex gets renormalized and the pairing eigenvalues in all channels decrease monotonically.
From the pure electronic effective interactions $V_{\text{EI}}$, an $f_{x^3-3xy^2}$-wave ($B_{1u}$) and $p$-wave ($E_{1u}$) pairing dominate for weak interactions ($U<1.2$), as shown in the inset of the Fig.~\ref{fig_figure3}(b). With increasing interaction, the pairing eigenvalues of the $d$-wave, $i$-wave and $s$-wave increase rapidly and they become dominant after $U$ exceeding 1.2, promoted by the antiferromagnetic fluctuations within the unitcell (see SM). These states are almost degenerate approaching the critical interaction, consistent with previous results~\cite{wu_PRL_21}.  

The pairing eigenvalues from including both contributions from EPC and EI ($V_{\text{tot}}$) clearly show three regimes in Fig.~\ref{fig_figure3}(c): EPC dominated regime, crossover regime and EI dominated regime. In the first regime, $U<0.8$, the difference between $s$-wave and $d$-wave pairing eigenvalues is not big and superconductivity is not necessarily $s$-wave but might be $d$-wave, as $s$-wave paring will be significantly reduced by the Coulomb pseudopotental $\mu^*$, while the subleading $E_{2g}$ pairing is unaffected, leading to a value for $T_c$ of the order of a few K (see SM). In the crossover regime, $p$-wave and $f_{y^3-3yx^2}$-wave ($B_{2u}$) pairing becomes dominant for $0.9<U<1.3$, in contrast to the discussed pure EI case, and the $d$-wave pairing is slightly favored for $1.3<U<2$. Through comparing pairing eigenvalues in Fig.~\ref{fig_figure3}(a), (b) and (c), we can find that there are two main effect triggering this behavior: a) repulsive $V_{\text{EI}}$, whose magnitude increases with increasing $U$ and $V$, significantly suppresses the onsite $A_{1g}$ and NN $E_{2g}$ pairing from the EPC; b) the interaction-driven $E_{2g}$ pairing gets slightly enhanced by EPC for $U>1.3$. Moreover, the  electronic interaction suppresses the pairing in the spin-singlet channels faster than the triplet channels. Therefore, the dominant $p$-wave and $f_{y^3-3yx^2}$-wave pairing emerge in the crossover regime owing to the combined effect of EPC and EI.

The situation for $\mu =-2.08$ is quite different.
As shown in Fig.~\ref{fig_figure3}(d), the $s$-wave pairing is dramatically dominant in the EPC case compared to $E_{2g}$, $E_{1u}$, and $B_{1u}$, different from the case of $p$-type VHS, which is consistent with the smaller anisotropy of the EPC effective interaction [Fig.~\ref{fig_figure2}(g)]. From $V_{\text{EI}}$, the $p$-wave and $B_{1u}$-wave pairing are leading with close eigenvalues for weak interactions $U<1.7$, as shown in the inset of the Fig.~\ref{fig_figure3}(e). The $A_{2g}$ ($i$-wave) pairing is subleading at weak interactions but increases rapidly with increasing $U$ ($U>1$) and its pairing eigenvalue exceeds the $p$-wave state and becomes leading for $U>1.7$. The $s$- and $d$-wave states are subleading close to the critical interaction, distinct from the $p$-type VHS. These states are promoted by the antiferromagnetic spin and bond fluctuations (see SM). Similarly, there are three regimes in Fig.~\ref{fig_figure3}(f) when including both EPC and EI. For $U<0.9$, an $s$-wave pairing from EPC dominates. In the crossover regime, the $p$-wave pairing is dominant for $0.9<U<1.1$ and $A_{2g}$-wave pairing becomes clearly leading for $1.1<U<1.8$, which is ascribed to the competition between EPC and EI in the $p$-wave channel and mutual enhancement in the $A_{2g}$ channel.

\textit{Conclusions and implications for experiments.--}
We have shown that the combined effect of electron-phonon coupling and electronic correlations near VH fillings is crucial for the occurrence of anomalous pairing in kagome materials. Although our results are for the Holstein-Hubbard model, they can be expected to hold for a more general class of models due to the intriguing sublattice texture.
Our theory can be applicable to recently discovered kagome superconductors, including AV$_3$Sb$_5$~\cite{AV3Sb5_Ortiz_first_paper,PhysRevLett.125.247002}, RERu$_3$Si$_2$ ~\cite{BARZ19801489,PhysRevMaterials.5.034803,ChinPhysLett.39.087401}, Ti-based materials~\cite{YangHT2022}
 and Ta$_2$V$_{3.1}$Si$_{0.9}$~\cite{LiuHX2023}.
Especially, for the 
multi-orbital AV$_3$Sb$_5$, where two $p$-type VHSs and one $m$-type VHS appear near the Fermi level~\cite{MKang2022,YHu2022}, our work has implications for its pairing symmetry. First, it is noted that the pairing eigenvalue $\lambda_s$ without correlations ($U=V=0$) is of the order of the values reported in first-principle calculations~\cite{Tan21,Zhang21,Si22,2022arXiv220702407Z}, confirming that our choice of phonon parameters is realistic for AV$_3$Sb$_5$. As nematicity emerges in the CDW order with the lattice rotational symmetry broken, the corresponding two-fold pairing states may be splitted. If AV$_3$Sb$_5$ is located at the crossover regime of EPC and EI, a nematic pairing, involving a mixture of $p_x$- and $p_y$-wave gaps, or an $f$-wave pairing, will be favored from the $p$-type VHS at ambient pressure. This gap is usually nodal or has a deep minimum, consistent with nodal signatures in experiments~\cite{2021arXiv210208356Z,2022arXiv220207713G}. With increasing pressure, the band width increases and the correlation effect weakens. Thus, the pairing may stay unchanged or transforms to an $s$-wave state according to Fig.\ref{fig_figure3}(c). When the CDW order is eliminated, the lattice rotational symmetry is restored and a $p+ip$ in the $E_{1u}$ channel with time-reversal symmetry breaking will be favored. Both $p+ip$- and $s$-wave states are nodeless. This provides a possible explanation for the observed transition from a nodal to nodeless gap with increasing pressure in $\mu$SR measurements~ \cite{2022arXiv220207713G}. Moreover, the $p+ip$ state can account for the  time-reversal symmetry breaking of the superconducting state observed in $\mu$SR measurements without CDW order.

\textit{Acknowledgements.--}
A.G. thanks the Max-Planck-Institute for Solid State Research in Stuttgart for hospitality and financial support. D.C. thanks Max-Planck-Institute for Solid State Research in Stuttgart for hospitality, and acknowledges financial support from Kungl. Vetenskapsakademien and C.F. Liljewalchs stipendiestiftelse Foundation. X.W. is supported by the National Key R\&D Program of China (Grant No.
2023YFA1407300) and the National Natural Science Foundation of China (Grant No. 12047503).
A.P.S. is funded by the Deutsche Forschungsgemeinschaft (DFG, German Research Foundation) – TRR 360 – 492547816. Numerical calculations in this work were performed on the HPC Cluster of ITP-CAS.

\textit{Note added.--}
Upon finalizing our manuscript, we became aware of Ref.~\cite{Romer2022}, where pure electron-electron mediated pairing was discussed on the Kagome lattice by including both bubble and ladder diagrams with onsite and nearest-neighbor interactions. A rich pairing phase diagram is revealed, in agreement with our findings.

\bibliography{references_newR}

\appendix

\section{ Model and methods}
\label{sm_sec1}

\subsection{Hamiltonian}

We consider a Hamiltonian on a two-dimensional kagome lattice with both electron-electron and electron-phonon interactions,
\begin{equation}
    \mathcal{H}=\mathcal{H}_e+\mathcal{H}_p+\mathcal{H}_{ep},
    \label{eq:totHamil}
\end{equation}
with the electron part of the Hamiltonian $H_e$ given by an extended Hubbard model,
\begin{align}
   &\mathcal{H}_e=\mathcal{H}_{0}+\mathcal{H}_{\rm int}, \nonumber \\
   &\mathcal{H}_0=-t\sum_{\substack{\langle ij \rangle,\alpha,\beta,\sigma \\ \alpha \neq \beta}} c^{\dagger}_{i\alpha\sigma}c_{j\beta\sigma} - \mu \sum_{i,\alpha} n_{i\alpha} \nonumber \\
   &\mathcal{H}_{\rm int}=U\sum_{i,\alpha} n_{i\alpha\uparrow}n_{i\alpha\downarrow} + V\sum_{\substack{\langle ij \rangle,\alpha,\beta \\ \alpha \neq \beta}} n_{i\alpha} n_{j\beta},
   \label{eq:elecHamil}
\end{align}
where $c^{\dagger}_{i\alpha\sigma}$ ($c_{i\alpha\sigma}$) is the creation (annihilation) operator of an electron with spin $\sigma$ at lattice site $i$, $\alpha=A,B,C$ is the index for the three different sublattices, $n_{i\alpha}=n_{i\alpha\uparrow}+n_{i\alpha\downarrow}$ is the electron density operator, $\langle ij \rangle$ denotes the nearest neighbor sites, $t$ is the nearest neighbor hopping amplitude, $\mu$ is the chemical potential, $U$ is the onsite Hubbard repulsion, and $V$ is the nearest neighbor Coulomb repulsion. For the phonon part of the Hamiltonian, we consider a Holstein phonon mimicking the out-of-plane vibration of ions at the centre of the kagome unit cell given by
\begin{equation}
   \mathcal{H}_{p}=\sum_{i} \hbar\omega_{D}\left(a_{i}^{\dagger}+a_{i}\right),
   \label{eq:phonHamil}
\end{equation}
where $a^{\dagger}_{i}$ ($a_{i}$) is the phonon creation (annihilation) operator and $\omega_D$ is the phonon frequency, which can be associated to the Debye frequency for studying superconductivity. The electron-phonon coupling is given as
\begin{equation}
   \mathcal{H}_{ep}=g_{0}\sum_{i,\alpha}n_{i\alpha}\left(a_{i}^{\dagger}+a_{i}\right),
   \label{eq:elecphonHamil}
\end{equation}
where $g_0$ is the bare electron-phonon coupling constant and within the considered model the electrons at different sublattice couple to the phonon with a same coupling strength.  For simplicity we express energies in units of $t$, lengths in units of the lattice spacing $a$, and set $\hbar=k_{B}=1$.

Due to the translational invariance, we Fourier transform the Hamiltonian using the transformation $c^{\dagger}_{i\alpha}=\frac{1}{\sqrt{N}}\sum_{k}e^{-\mathbf{k}\cdot\mathbf{R}_i}c^{\dagger}_{\mathbf{k}\alpha}$ for the electron operators and $a^{\dagger}_{i}=\frac{1}{\sqrt{N}}\sum_{q}e^{-\mathbf{q}\cdot\mathbf{R}_i}a^{\dagger}_{\mathbf{q}}$ for the phonon operators where $N$ is the total number of sites. The phonon and the electron-phonon part of the Hamiltonian can be written as
\begin{eqnarray}
    H_{p}+H_{ep}&=&\sum_{\mathbf{q}} \omega_{D}\left( a^{\dagger}_{\mathbf{q}}a_{\mathbf{q}} + \frac{1}{2}\right)\nonumber\\
    &+&\frac{g_0}{\sqrt{N}}\sum_{\alpha\mathbf{k}\mathbf{q}\sigma} c_{\mathbf{k}+\mathbf{q}\alpha\sigma}^{\dagger}c_{\mathbf{k}\alpha\sigma}\left(a_{-\mathbf{q}}^{\dagger}+a_{\mathbf{q}}\right).
    \label{eq:pepmom}
\end{eqnarray}
After Fourier transformation, and making explicit the spin indices, the non-interacting Hamiltonian can be written in a matrix form $H_0=\sum_{\mathbf{k}}\Phi^{\dagger}h_0(\mathbf{k})\Phi$ in the basis $\Phi^{\dagger}=\left( c^{\dagger}_{\mathbf{k} A \uparrow},c^{\dagger}_{\mathbf{k} A \downarrow},c^{\dagger}_{\mathbf{k} B \uparrow},c^{\dagger}_{\mathbf{k} B \downarrow},c^{\dagger}_{\mathbf{k} C \uparrow},c^{\dagger}_{\mathbf{k} C \downarrow} \right)$ with
\begin{equation}
    h_0(\mathbf{k})=\left( \begin{array}{ccc} -\mu I_{2} & -t f_{\rm AB}I_{2} & -t f_{AC}I_{2} \\ -tf_{\rm BA}I_{2} & -\mu I_{2} & -tf_{\rm BC}I_{2} \\ -t f_{\rm CA}I_{2} & -t f_{\rm CB}I_{2} & -\mu I_{2} \end{array} \right),
    \label{eq:H0matrix}
\end{equation}
where $I_{2}$ is a $2\times2$ identity matrix in the spin basis, the elements are given by
\begin{align}
    &f_{AB}(\mathbf{k})=\left(1+e^{-i\mathbf{k}\cdot\mathbf{a}_1}\right), \nonumber \\
    &f_{AC}(\mathbf{k})=\left(1+e^{-i\mathbf{k}\cdot\left(\mathbf{a}_1+\mathbf{a}_2\right)}\right), \nonumber \\
    &f_{BC}(\mathbf{k})=\left(1+e^{-i\mathbf{k}\cdot\mathbf{a}_2}\right),
    \label{eq:intmatrix2}
\end{align}
and $f_{\alpha\beta}(\mathbf{k})=f_{\beta\alpha}(-\mathbf{k})$, with the unit vectors given as $\mathbf{a}_{1}=\left(\sqrt{3}/2,-1/2\right)^{T}$ and $\mathbf{a}_{2}=\left(0,1\right)^{T}$. The corresponding non-interacting electronic Green's function is obtained by ${G^{0}}(k,i\omega_n)=(i\omega_n-h_0(\mathbf{k}))^{-1}$.

\subsection{ Interactions within Random Phase Approximation}

\begin{figure}[t!]
\begin{center}
\includegraphics[width=0.48\textwidth]{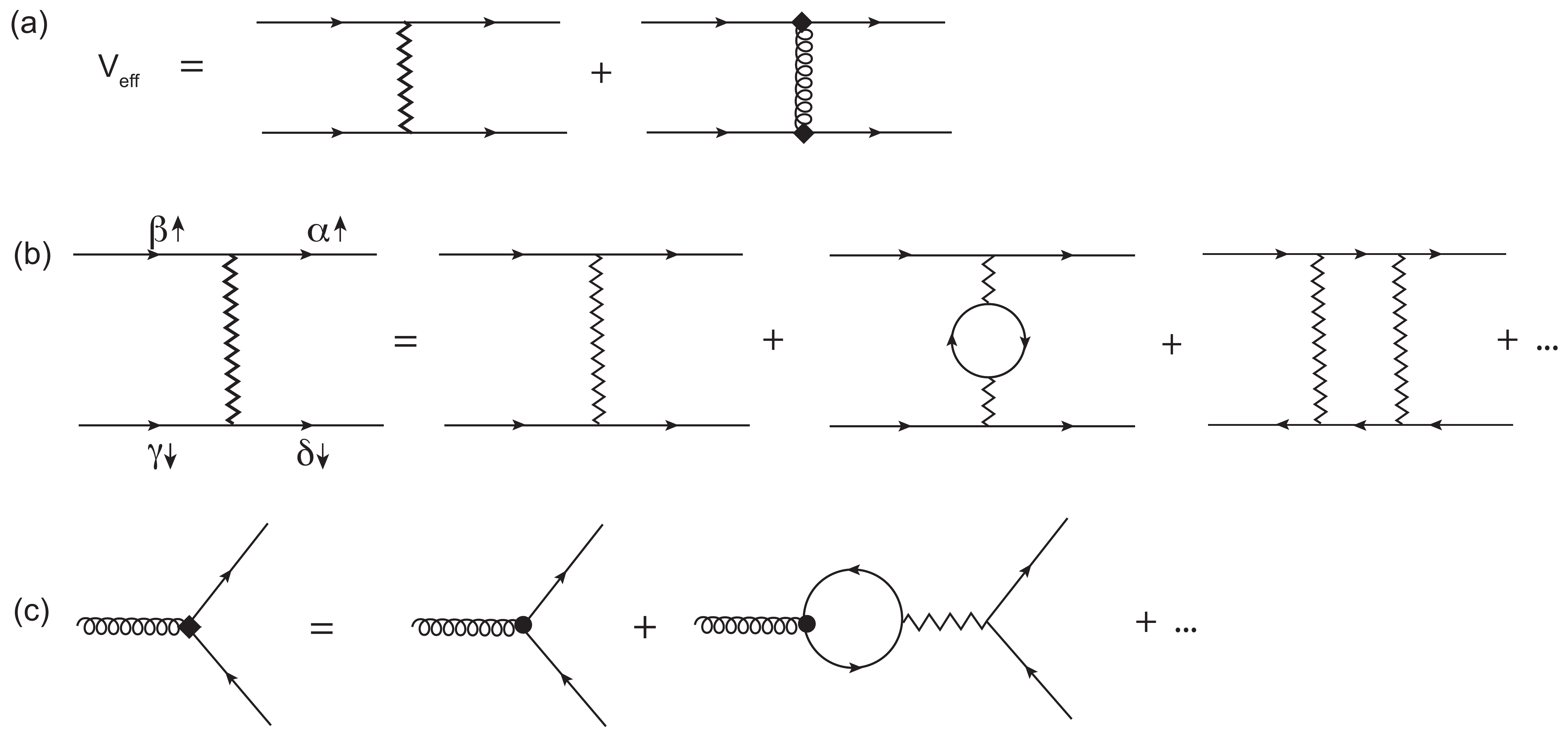}
\caption{(a) Total effective interaction  $\tilde{V}_{eff}^{\rm tot}$ from electronic and EPC interactions. (b) Bubble and Ladder diagrams for electronic interactions. (c) The renormalized EPC vertex. Solid zig-zag line represent the effective electron-electron interaction, filled diamonds indicate the renormalized electron-phonon interaction vertex of the bare electron-phonon interaction vertex denoted as filled circle, curly line indicate the phonon propagator, and the solid lines represent electron propagator.}
\label{fig_rpa}
\end{center}
\end{figure}

The total effective interaction can be written as the sum of the electron-electron (ee) and EPC interactions (Fig.~\ref{fig_rpa} (a)).
From the bubble diagrams, we calculate the effective electron-electron (ee) interactions incorporating the renormalizations caused due to charge and spin fluctuations as shown in Fig.~\ref{fig_rpa} (b),
\begin{equation}
    \{\tilde{V}_{eff}^{\rm ee,B}\left(\mathbf{q},i\nu_n\right)\}^{\sigma\sigma'}_{\alpha\alpha,\beta\beta}=\{\left[I+V_{\rm int}^{0}\left(\mathbf{q}\right) \chi^{0}\left(\mathbf{q},i\nu_n\right) \right]^{-1}V_{\rm int}^{0}\left(\mathbf{q}\right)\}^{\sigma\sigma'}_{\alpha,\beta},
    \label{eq:effecintsubel}
\end{equation}
where $I$ is a $6\times6$ identity matrix, the bare (non-interacting) susceptibility $\chi^{0}$ is given by
\begin{equation}
    \chi_{\alpha\sigma,\beta\sigma^\prime}^{0}\left(\mathbf{q},i\nu_n\right)=-\frac{T}{N}\sum_{k,i\omega_m}G_{\alpha\sigma,\beta\sigma^\prime}^{0}(\mathbf{k},i\omega_m)G_{\beta\sigma^\prime,\alpha\sigma}^{0}(\mathbf{k}+\mathbf{q},i\omega_n+i\nu_n),
    \label{eq:baresusc}
\end{equation}
where $i\omega_n$ ($i\nu_n$) is the fermionic (bosonic) matsubara frequency, $T$ is the temperature, and $G_{\alpha\sigma\beta\sigma^\prime}^{0}(k,i\omega_n)$ are elements of the non-interacting electronic Green's function ${G^{0}}(k,i\omega_n)$. The bare interaction matrix is defined as
\begin{align}
    &{[V^{0}_{\rm int}]}_{\alpha\sigma,\beta\sigma^\prime}(\mathbf{q})=U\left(1-\delta_{\sigma\sigma^\prime}\right)\delta_{\alpha\beta}+Vf_{\alpha\beta}\left(1-\delta_{\alpha\beta}\right),
    \label{eq:intmatrix1}
\end{align}
where $\delta$ is the kronecker delta function and $f_{\alpha\beta}$ are given by Eq.~\eqref{eq:intmatrix2}.

In the ladder diagrams, the interaction lines dependends on the internal momenta and the geometric summation cannot be directly done with previously defined susceptibilities. In order to address it, we introduce a generalized susceptibility~\cite{PhysRevB.104.064507},
\begin{eqnarray}
[\chi^L_0(\bm{q},i\nu_n)]^{st}_{\alpha\beta,\gamma\delta}&=&-\frac{T}{N}\sum_{\bm{k},i\omega_m}f^{s,*}_{\alpha\beta}(\bm{k})G^0_{\alpha\gamma}(\bm{k}) \nonumber\\
&\times& G^0_{\delta\beta}(\bm{k}+\bm{q},i\omega_m+i\nu_n)f^{t}_{\gamma\delta}(\bm{k}), \nonumber\\
\end{eqnarray}
where $\alpha,\beta,\gamma,\delta$ is the sublattice index and $s,t=0,+,-$ denotes the onsite, symmetrical and antisymmetric nearest-neighbor form factors. The corresponding form factors are given by,
\begin{eqnarray}
   f^0_{\alpha\alpha}(\mathbf{k})&=&1, \nonumber\\
    f^{\pm}_{AB}(\mathbf{k})&=&1\pm e^{-i\mathbf{k}\cdot\mathbf{a}_1}, \nonumber \\
    f^{\pm}_{AC}(\mathbf{k})&=&1\pm e^{-i\mathbf{k}\cdot\left(\mathbf{a}_1+\mathbf{a}_2\right)}, \nonumber \\
    f^{\pm}_{BC}(\mathbf{k})&=&1\pm e^{-i\mathbf{k}\cdot\mathbf{a}_2}, \nonumber\\
    f^{\pm}_{\alpha\beta}(\mathbf{k})&=&f^{\pm}_{\beta\alpha}(-\mathbf{k}).
    \label{formfactor}
\end{eqnarray}
The general susceptibility is motivated by the separation of momenta in the interaction form factor, i.e. $f^+_{\alpha\beta}(-\bm{k}+\bm{k}')=\frac{1}{2}[f^{+,*}_{\alpha\beta}(\bm{k})f^{+}_{\alpha\beta}(\bm{k}')+f^{-,*}_{\alpha\beta}(\bm{k})f^{-}_{\alpha\beta}(\bm{k}') ] $. From the ladder diagrams, the renormalized susceptibility reads,
\begin{eqnarray}
[\chi^L_{RPA}(\bm{q},0)]^{st}_{\alpha\beta,\gamma\delta}=\{\chi^L_{0}(\bm{q},0)[1-\mathcal{V}_0\chi^L_{0}(\bm{q},0)]^{-1}\}^{st}_{\alpha\beta,\gamma\delta},\nonumber\\
\end{eqnarray}
with the  interaction matrix $\mathcal{V}_0$,
\begin{eqnarray}
[\mathcal{V}_0]^{s}_{\alpha\beta,\gamma\delta}=
\begin{cases}
U & \alpha=\beta=\gamma=\delta, s=0 \\
\frac{1}{2}V &  \alpha=\gamma, \beta=\delta, \alpha\neq \beta, s=\pm \\
0 &   \text{others} \\
\end{cases}
\end{eqnarray}
The effective Cooper pair scattering through the ladders diagram reads,
\begin{eqnarray}
[\tilde{V}^{ee,L}_{sc}(\bm{k},\bm{k}')]^{\uparrow\downarrow}_{\alpha\beta,\gamma\delta}=[\mathcal{V}_{\bm{k}}\chi^L_{RPA}(\bm{k}+\bm{k}',0)\mathcal{V}_{\bm{k}'}]_{\delta\beta,\gamma\alpha},
\end{eqnarray}
with the interaction matrix $\mathcal{V}_{\bm{k}}$,
\begin{eqnarray}
[\mathcal{V}_{\bm{k}}]^{s}_{\alpha\beta,\gamma\delta} =
\begin{cases}
U & \alpha=\beta=\gamma=\delta, s=0 \\
\frac{1}{2}V f^s_{\alpha\beta}(\bm{k}) &  \alpha=\gamma, \beta=\delta, \alpha\neq \beta, s=\pm \\
0 &   \text{others} \\
\end{cases}
\nonumber\\
\end{eqnarray}

For the electron-phonon effective interaction the second diagram of Fig.~\ref{fig_rpa} (a) is computed in RPA~\cite{schrieffer1964theory,mahan2000many}, where the bare electron-phonon coupling constant $g_0$ is renormalized due to charge fluctuations, represented by the diagram in Fig.~\ref{fig_rpa} (c), giving
\begin{equation}
    \tilde{g}\left(\mathbf{q},i\nu_n\right)=\left[I+ V_{\rm int}^{0}\left(\mathbf{q}\right) \chi^{0}\left(\mathbf{q},i\nu_n\right) \right]^{-1}\hat{g}_{0},
    \label{eq:gtilde}
\end{equation}
where $\hat{g}_{0}=g_0I_{6\times1}$ and $I_{6\times1}$ is a $6\times1$ matrix with all components being unity.

\subsection{Effective pairing interactions and superconducting instabilities}

We investigate the possible superconducting instabilities
induced by both spin and charge fluctuations. Fluctuation-induced renormalized effective electron-electron interaction
can give rise to unconventional superconductivity as often discussed in the context of high $T_c$ superconductors \cite{Scalapino86,Onari04,Graser09}. Importantly, the renormalized electron-phonon coupling constant in Eq.~\eqref{eq:gtilde} also gives the possibility of the emergence of unconventional supercoducting pairings \cite{Foussats_PRB_05,Foussats_2006}. The total effective interaction in the superconducting channel in the weak coupling BCS form is
\begin{equation}
    H_{\rm sc}^{eff}=\sum_{\alpha,\beta,\mathbf{k},\mathbf{k}^\prime}\left[\tilde{V}_{\rm sc}^{\rm tot}\left(\mathbf{k},\mathbf{k}^{\prime}\right)\right]_{\alpha\beta} c^{\dagger}_{\alpha\mathbf{k}\uparrow}c^{\dagger}_{\beta-\mathbf{k}\downarrow}c_{\beta-\mathbf{k}^\prime\downarrow}c_{\alpha\mathbf{k}^{\prime}\uparrow},
    \label{eq:effecintsub}
\end{equation}

\noindent where $\tilde{V}_{\rm sc}^{\rm tot}\left(\mathbf{k},\mathbf{k}^{\prime}\right)$ is the total pairing scattering vertex in the sublattice basis and is a sum of the pure electronic contribution $\tilde{V}_{\rm sc}^{\rm ee,B}\left(\mathbf{k},\mathbf{k}^{\prime}\right)$, $\tilde{V}_{\rm sc}^{\rm ee,L}\left(\mathbf{k},\mathbf{k}^{\prime}\right)$ and the pure electron-phonon contribution $\tilde{V}_{\rm sc}^{\rm ep}\left(\mathbf{k},\mathbf{k}^{\prime}\right)$, also see Fig.~\ref{fig_rpa}(a). The paure electronic contribution from both bubbles and ladders in the $\uparrow\downarrow$ channel is given as
\begin{widetext}
\begin{equation}
    \tilde{V}_{\rm sc}^{\rm ee,\uparrow\downarrow}\left(\mathbf{k},\mathbf{k}^{\prime}\right)=\left( \begin{array}{ccc} \left[\tilde{V}_{eff}^{\rm ee,B}\left(\mathbf{q}\right)\right]_{12} & \left[\tilde{V}_{eff}^{\rm ee,B}\left(\mathbf{q}\right)\right]_{14} & \left[\tilde{V}_{eff}^{\rm ee,B}\left(\mathbf{q}\right)\right]_{16} \\ \left[\tilde{V}_{eff}^{\rm ee,B}\left(\mathbf{q}\right)\right]_{32} & \left[\tilde{V}_{eff}^{\rm ee,B}\left(\mathbf{q}\right)\right]_{34} & \left[\tilde{V}_{eff}^{\rm ee,B}\left(\mathbf{q}\right)\right]_{36} \\ \left[\tilde{V}_{eff}^{\rm ee,B}\left(\mathbf{q}\right)\right]_{52} & \left[\tilde{V}_{eff}^{\rm ee,B}\left(\mathbf{q}\right)\right]_{54} & \left[\tilde{V}_{eff}^{\rm ee,B}\left(\mathbf{q}\right)\right]_{56} \end{array} \right)+[\tilde{V}^{ee,L}_{sc}(\bm{k},\bm{k}')]^{\uparrow\downarrow},
    \label{eq:eeSCint}
\end{equation}
\end{widetext}

Due to the absence of any spin orbit interactions, we only keep the opposite spin terms of $\tilde{V}_{eff}^{\rm ee}\left(\mathbf{q}\right)$ making $\tilde{V}_{\rm sc}^{\rm ee}\left(\mathbf{k},\mathbf{k}^{\prime}\right)$ a $3\times3$ matrix in the sublattice basis. The pure electron-phonon contribution is represented by the second diagram in the right hand side of Fig.~\ref{fig_rpa} (a) giving
\begin{equation}
    \tilde{V}_{\rm sc}^{\rm ep}\left(\mathbf{k},\mathbf{k}^{\prime}\right)=D(\mathbf{q},i\nu_n=0)\left( \begin{array}{ccc} \tilde{g}_1\tilde{g}_2 & \tilde{g}_1\tilde{g}_4 & \tilde{g}_1\tilde{g}_6 \\ \tilde{g}_3\tilde{g}_2 & \tilde{g}_3\tilde{g}_4 & \tilde{g}_3\tilde{g}_6 \\ \tilde{g}_5\tilde{g}_2 & \tilde{g}_5\tilde{g}_4 & \tilde{g}_5\tilde{g}_6 \end{array} \right),
    \label{eq:epSCint}
\end{equation}
where $D(\mathbf{q},i\nu_n)=2\omega_{D}/((i\nu_n)^{2}-\omega_{D}^2)$ is the bare phonon propagator. Here again, the absence of spin-orbit coupling enables us to write $\tilde{V}_{\rm sc}^{\rm ep}\left(\mathbf{k},\mathbf{k}^{\prime}\right)$ as a $3\times3$ matrix since $\tilde{g}_1=\tilde{g}_2$, $\tilde{g}_3=\tilde{g}_4$ and $\tilde{g}_5=\tilde{g}_6$.
Here $\mathbf{q}=\mathbf{k}-\mathbf{k}^\prime$
 and we take, as usual~\cite{wu_PRL_21,roemer_2022_arXiv,Foussats_PRB_05,Motrunich04}$, \mathbf{k}$ and $\mathbf{k}^\prime$ to be restricted to the Fermi surface and $i\nu_n=0$ in the spirit of a weak coupling BCS approach, i.e., no retardation effects are included.

Now, we transform the effective interaction in Eq.~\eqref{eq:effecintsub} in the band basis using the transformation that diagonalizes the non-interacting Hamiltonian $h_0(\mathbf{k})$,
\begin{equation}
    c^{\dagger}_{\mathbf{k} \alpha \sigma}=\sum_{\gamma} \psi^{\dagger}_{\mathbf{k}\gamma}a^{*}_{\gamma\alpha\sigma},
    \label{eq:subtoband}
\end{equation}
where $\alpha$ is the sublattice index and $\gamma$ is the band index. Three sublattices in a Kagome lattice result into three bands. However, since we are interested in the effective interactions in the superconducting channel within weak coupling formalism, we only consider the band crossing the Fermi energy. As a result, each $\mathbf{k}$ corresponds to only one band and hence we drop the index $\gamma$ hereafter. Using the transformation in Eq.~\eqref{eq:subtoband}, Eq.~\eqref{eq:effecintsub} can be written in the band basis as
\begin{equation}
    H_{\rm sc}^{eff}=\sum_{\mathbf{k}\mathbf{k}^\prime}V_{\rm tot}\left(\mathbf{k},\mathbf{k}^\prime\right) \psi^{\dagger}_{\mathbf{k}\uparrow}\psi^{\dagger}_{\mathbf{-k}\downarrow}\psi_{-\mathbf{k}^\prime\downarrow}\psi_{\mathbf{k}^\prime\uparrow},
    \label{eq:effecintband}
\end{equation}
where $V_{\rm tot}\left(\mathbf{k},\mathbf{k}^\prime\right)=V_{\rm EI}\left(\mathbf{k},\mathbf{k}^\prime\right)+V_{\rm EPC}\left(\mathbf{k},\mathbf{k}^\prime\right)$ with
\begin{eqnarray}
    V_{\rm EPC/EI}\left(\mathbf{k},\mathbf{k}^\prime\right)&=&\sum_{\alpha\beta}a^{*}_{\alpha\uparrow}(\mathbf{k})a^{*}_{\beta\downarrow}(-\mathbf{k})a_{\beta\downarrow}(-\mathbf{k}^{\prime})\nonumber\\
    &\times&a_{\alpha\uparrow}(\mathbf{k}^{\prime})\left[\tilde{V}^{ep/ee}\left(\mathbf{k},\mathbf{k}^\prime\right)\right]_{\alpha\beta},
    \label{eq:projectoband}
\end{eqnarray}
where $\tilde{V}^{ep/ee}\left(\mathbf{k},\mathbf{k}^\prime\right)$ are given by Eqs.~\eqref{eq:eeSCint} and \eqref{eq:epSCint} in the sublattice space.

We compare different superconducting instabilities using the effective interactions $V_{\rm EPC/EI}$ and $V_{\rm tot}$. We apply two approaches. The first approach is to compute the effective sperconducting couplings using the eigenvalues of the matrix $V_{\rm EPC/EI/\rm {tot}}$ in the singlet (S) and the triplet (T) channels separately using
\begin{equation}
    V^{\rm{S}/\rm{T}}_{\rm{EPC}/\rm{EI}/\rm {tot}}\left(\mathbf{k},\mathbf{k}^\prime\right)=\frac{1}{2}\left[ V_{\rm{EPC}/\rm{EI}/\rm{tot}}\left(\mathbf{k},\mathbf{k}^\prime\right) \pm V_{\rm{EPC}/\rm{EI}/\rm{tot}}\left(\mathbf{k},-\mathbf{k}^\prime\right) \right],
    \label{eq:singtrip}
\end{equation}
which is equivalent to finding the eigenvalue of the linearized gap equation,
\begin{equation}
  -\int_{\rm FS} \frac{\sqrt{3}d {\bf k'}}{2(2 \pi)^2|v_{\bf k'}|}V^{\rm{S}/\rm{T}}_{\rm{EPC}/\rm{EI}/\rm {tot}}\left(\mathbf{k},\mathbf{k}^\prime\right) \Delta_{i}(\mathbf{k}^{\prime})=\lambda_i^ {\rm{EPC}/\rm{EI}/\rm {tot}}\Delta_{i}(\mathbf{k}),
  \label{eq:gapeigenvalues}
\end{equation}
where $\lambda_i^{\rm{EPC}/\rm{EI}/\rm {tot}}$ are the eigenvalues corresponding to the gap function $\Delta_{i}(\mathbf{k})$ and $v_{\mathbf k}$ is the quasiparticle velocity at momentum ${\mathbf k}$.

Another approach involves computing the effective couplings $\lambda_i$ in the different pairing channels or irreducible representations of the order parameter on the kagome lattice after projecting the interactions on a particular pairing channel. $\lambda_i$ is then defined as
\begin{widetext}
\begin{eqnarray}
\lambda_i^{\rm{EPC}/\rm{EI}/\rm{tot}}=-\frac{\sqrt{3}}{2(2 \pi)^2}
\frac{\int_{\rm FS} (d {\bf k} /|v_{\bf k}|) \int_{\rm FS} (d {\bf k'}/|v_{\bf k'}|)
F_i({\bf k'})
V_{\rm{EPC}/\rm{EI}/\rm{tot}}\left(\mathbf{k},\mathbf{k}^\prime\right) F_i({\bf k})}{
\int_{\rm FS} (d {\bf k}/|v_{\bf k}|) F_i({\bf k})^2 }\, ,
\label{eq:lambdaiproj}
\end{eqnarray}
\end{widetext}
where $F_{i}(\mathbf{k})$ denotes the different pairing symmetry channels \cite{Motrunich04,wu_PRL_21}. $\lambda_i$ measures the strength of the interaction between electrons at the Fermi surface in a given symmetry channel $i$. Superconductivity is only possible when $\lambda_i > 0$. Note that in this method of computing the effective couplings we ignore any mixing between different pairing channels. In principle, the gap function $\Delta_{i}(\mathbf{k})$ is a linear combination of different pairing symmetry channels $F_{i}(\mathbf{k})$ and contains mixing of different pairing channels with the same symmetry. However, we call the effective couplings obtained from Eqs.~\eqref{eq:gapeigenvalues} and \eqref{eq:lambdaiproj} using two different approaches both as $\lambda_i$ for simplicity. Moreover, we find similar findings in the main text where we use the eigenvalue method and in Sec. S5 where we use projection method.

\section{Spin and charge susceptibilties }
\begin{figure}[tp]
\begin{center}
\includegraphics[width=0.49\textwidth]{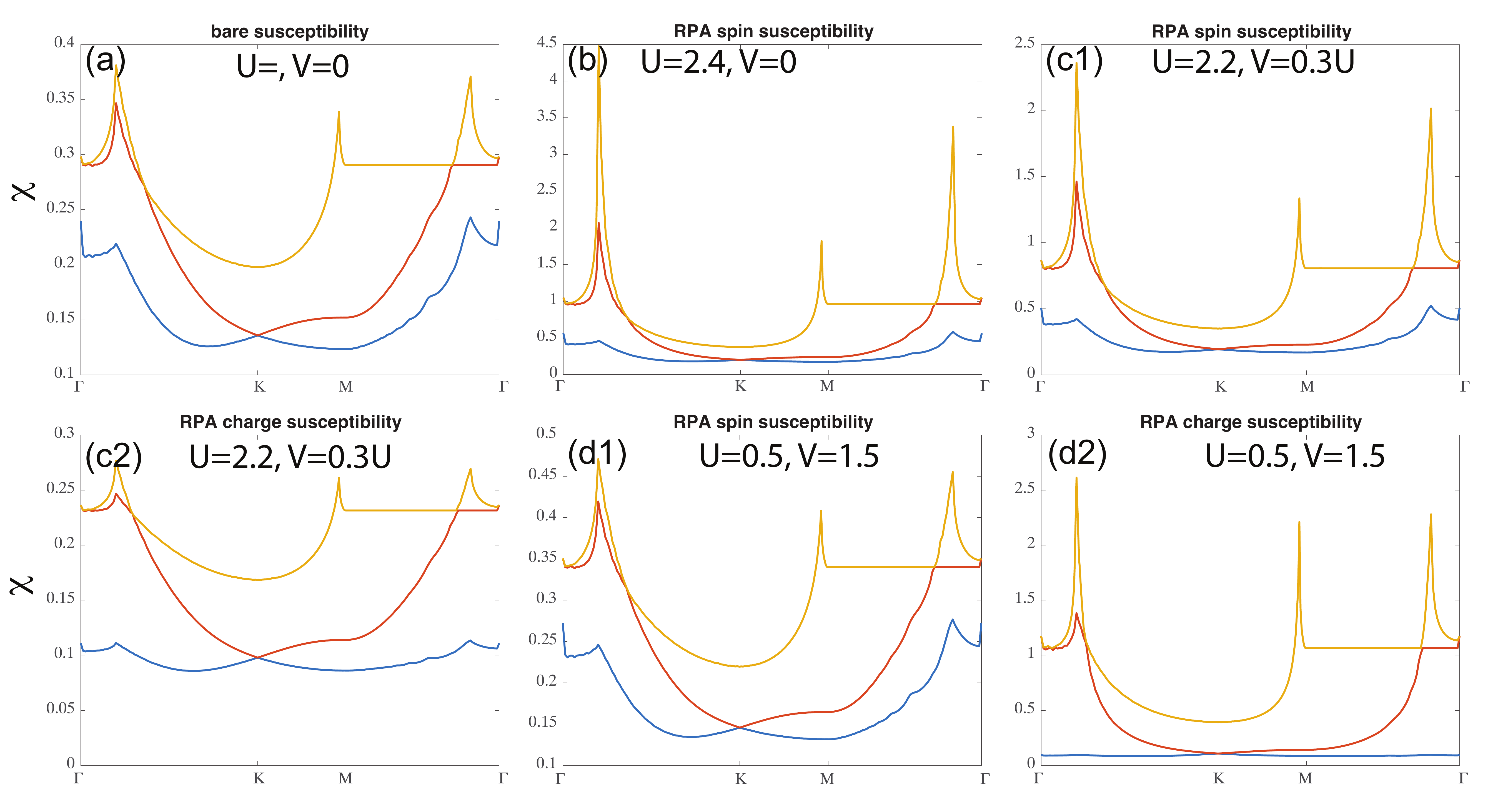}
\caption{ Eigenvalues of the spin and charge susceptibilities for different interaction parameters for $\mu=0.08$ near the p-type VHS filling.
}
\label{sus_ptype}
\end{center}
\end{figure}

Around the p-type VHS filling, the RPA spin and charge susceptibilities are shown in Fig.\ref{sus_ptype}. In the bare susceptibility, as discussed in the main text, there are prominent peaks around the $\Gamma$ and M points. An onsite interaction will significantly enhance the peak around the $\Gamma$ point and this peak corresponds to an antiferromagnetic fluctuation within the unitcell. The antiferromagnetic fluctuations usually promote spin-singlet pairing.  The nearest-neighbor repulsion will strongly enhance the peak around the M point in the charge channel. The divergence of RPA charge susceptibility corresponds to a $2\times2$ CDW instability. 

\begin{figure}[tp]
\begin{center}
\includegraphics[width=0.49\textwidth]{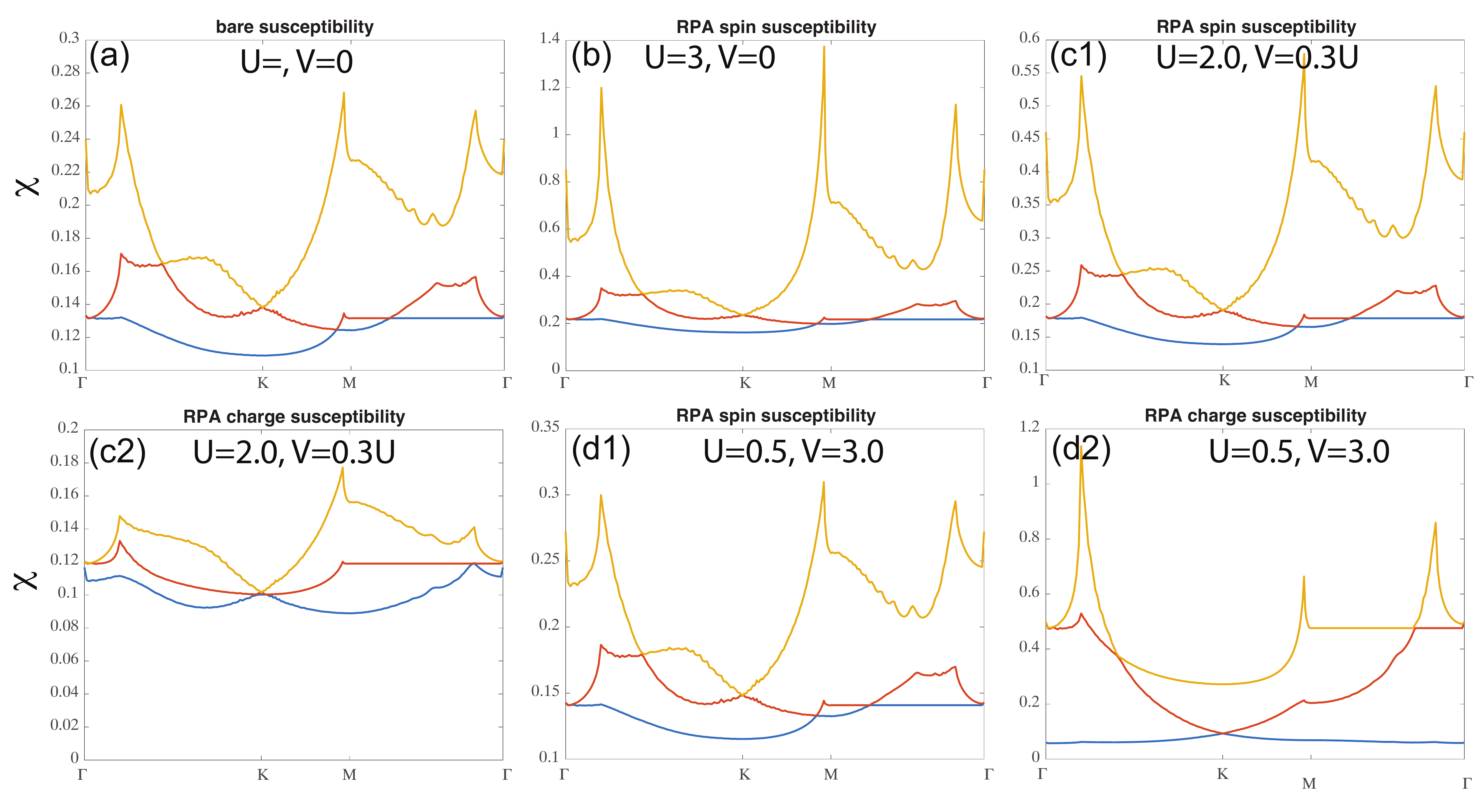}
\caption{ Eigenvalues of the spin and charge susceptibilities for different interaction parameters for $\mu=-2.08$ near the m-type VHS filling.
}
\label{sus_mtype}
\end{center}
\end{figure}

Around the m-type VHS filling, the RPA spin and charge susceptibilities are shown in Fig.\ref{sus_mtype}. Despite different sublattice texture on the Fermi surface, the prominent peaks for the m-type filling are similar to the p-type VHS for the bare case $U=V=0$. However, for the interacting case the susceptibilities are quite different.  An onsite interaction will significantly enhance the peak around the M point, corresponds to an antiferromagnetic fluctuation.   The nearest-neighbor repulsion will strongly enhance the peak around the $\Gamma$ point in the charge channel. 
 
\section{ Pairing eigenvalues with varying nearest-neighbor repulsion }
\begin{figure}[t]
\begin{center}
\includegraphics[width=0.49\textwidth]{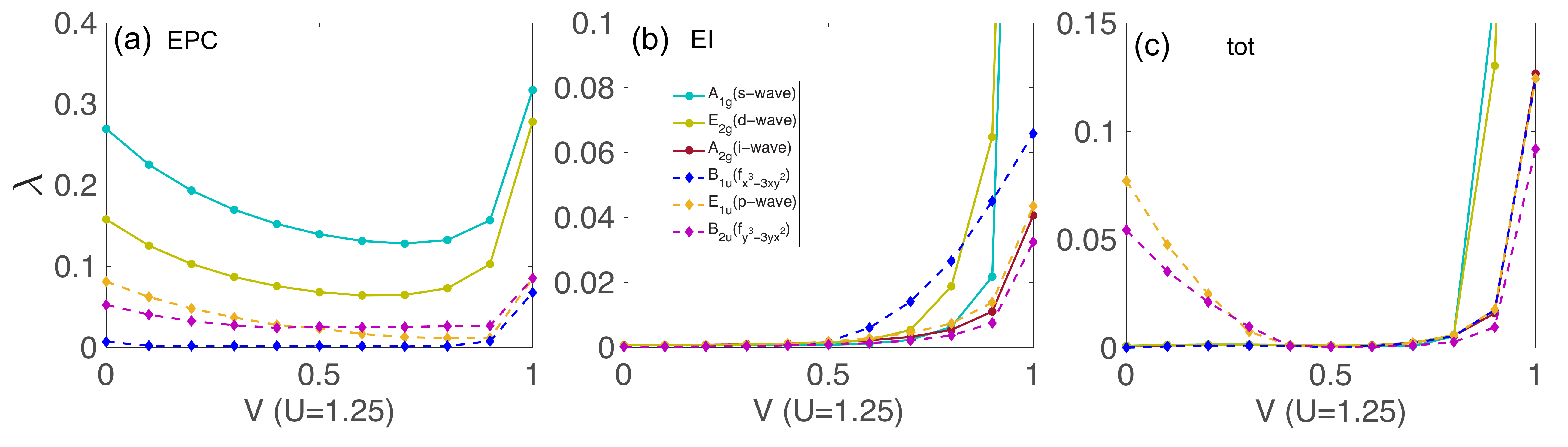}
\caption{Pairing eigenvalues for the leading pairing states as a function of $V$ ($U=1.25$) for $\mu=0.08$ near the p-type VHS filling: EPC (a), electronic interactions (b) and their summation (c).
}
\label{fig_V}
\end{center}
\end{figure}

In the main text, we study the pairing eigenvalues as a function of $U$ with a fixed ratio $V/U=0.3$. Here we further study the effect of tuning the nearest-neighbor repulsion. Fig.\ref{fig_V} displays the pairing eigenvalues for the leading pairing states as a function of $V$ in the crossover regime at $U=1.25$ near the $p$-type VHS filling. For the EPC the pairing eigenvalues gradually decrease with increasing $V$ and then increase abruptly from $V=0.9$, where $s$-wave pairing is always leading. In contrast, for the EI, the pairing eigenvalues increase with increasing $V$ and spin-triplet pairing dominant for $V<0.8$. When $V$ further increases, the pairing eigenvalues of both $s$-wave and $d$-wave states increase abruptly and become the leading ones. With including both EPC and EI, the dominant pairings are spin-triplet for weak NN repulsion and the pairing eigenvalues decrease to almost zeros with increasing $V$ towards $V=0.5$.  With further increasing $V$ towards the critical value $V_c \sim 1.1$ where the charge instability occurs, $s$-wave and $d$-wave pairings get enhanced abruptly and become dominant. Interestingly, it was suggested that pressure would increase $V$ driving the system into the superconducting state (see for instance Ref.[\onlinecite{merino01}]).

\section{Pairing instabilities away from the van Hove fillings}

\begin{figure}[h!]
\begin{center}
\includegraphics[width=0.49\textwidth]{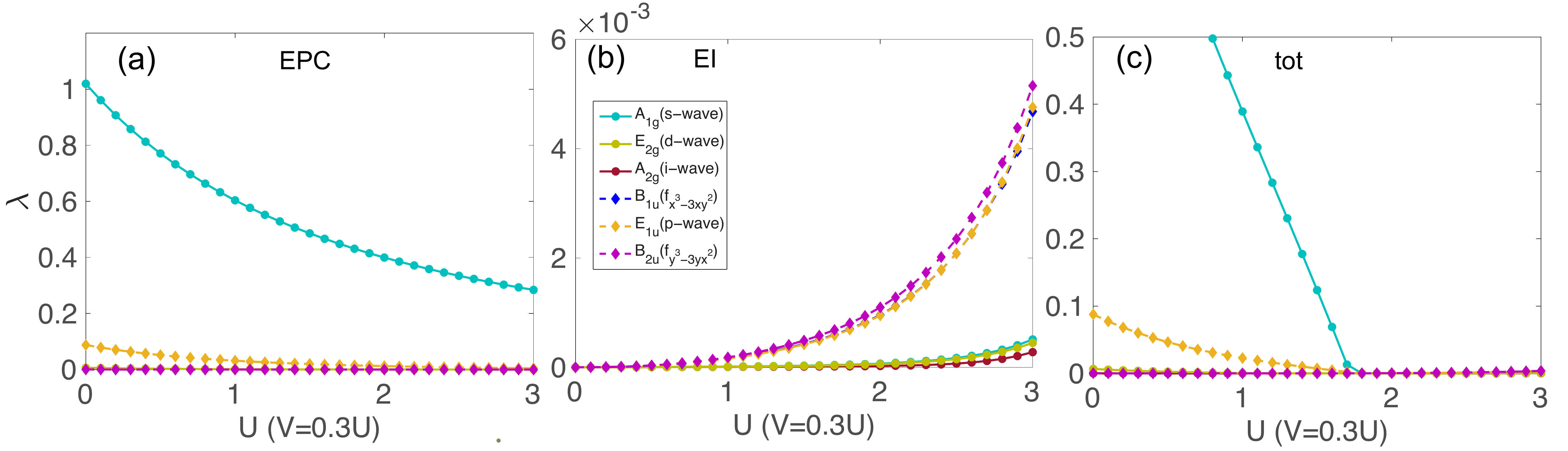}
\caption{
Pairing eigenvalues for the dominant instabilities as a function of U (V=0.3U) for $\mu=-3.5$ away from the VHS filling. The contribution from EPC, EI, electronic interactions, and the total case and in (a), (b), and (c), respectively.
}
\label{fig_figureS1}
\end{center}
\end{figure}

Away from the van Hove fillings, the sublattice distribution on the Fermi surface is uniform and the sublattice interference effect is very weak. As shown in Fig.~\ref{fig_figureS1} (a), EPC generate a dominant pairing in the isotropic $s$-wave $A_{1g}$ channel and the pairing eigenvalues of other spin-singlet pairings are extremely weak, similar to the EPC on the square and triangulat lattices [S4,S5]. In the spin-triplet channel,  there is only a weak  subleading $E_{1u}$ pairing at low $U$. The contributions from EI (Fig.~\ref{fig_figureS1} (b)) are very small, with $\lambda_i$'s of the order of $\sim 10^{-3}$. Finally, the total contribution (Fig.~\ref{fig_figureS1} (c)) is dominated the  $s$-wave channel from phonons. It is interesting to note that no superconductivity is expected for $U \gtrsim 1.8$, where the repulsive $s$-wave contributions from EI  suppresses the superconducting paring in $A_{1g}$ from phonons.

\section{ Pairing states with only bubble diagrams and Results of the projection} \label{proj}
In this section, and for completeness, we discuss the pairing with only bubble diagrams in the electronic interactions.

\begin{figure}[h!]
\begin{center}
\includegraphics[width=0.49\textwidth]{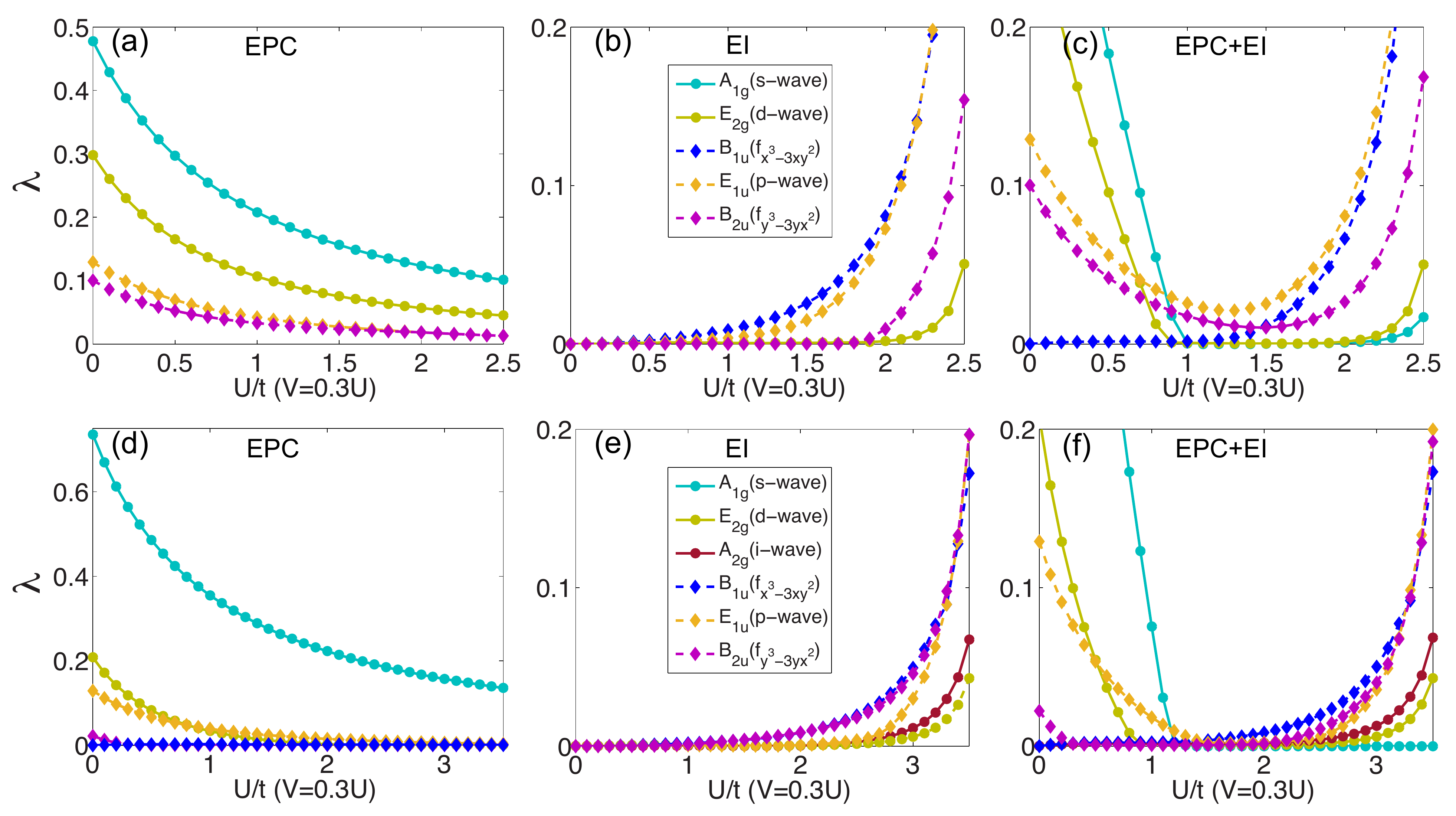}
\caption{
Pairing strength eigenvalues for the dominant instabilities as a function of U (V=0.3U) near the p-type (top panels) and m-type (bottom panels) and the contribution from EPC (a,d), electronic interactions (b,e) and their summation (c,f). Here only the bubble contributions are included. The eigenvalues are scaled by a factor of $2/\sqrt{3}$.
}
\label{fig_figure_bubbles}
\end{center}
\end{figure}

The pairing eigenvalues from EPC, EI and their summation by solving the the linearized gap equation are displayed in Fig.\ref{fig_figure_bubbles} (the eigenvalues are scaled by a factor of $2/\sqrt{3}$ compared with those in the main text). In contrast to Fig.3 in the main text, we find that the effective interactions from the bubble diagrams mainly promote the spin-triplet pairing for $U>1$. Similar to the main text, the electronic interactions will significantly suppress spin-singlet pairing from EPC at both van Hove fillings (see Fig.\ref{fig_figure_bubbles}(c),(f)) for large $U$. However, for $U<1$ EPC dominate with and without ladders contributions.

\begin{figure}[b]
\begin{center}
\includegraphics[width=0.48\textwidth]{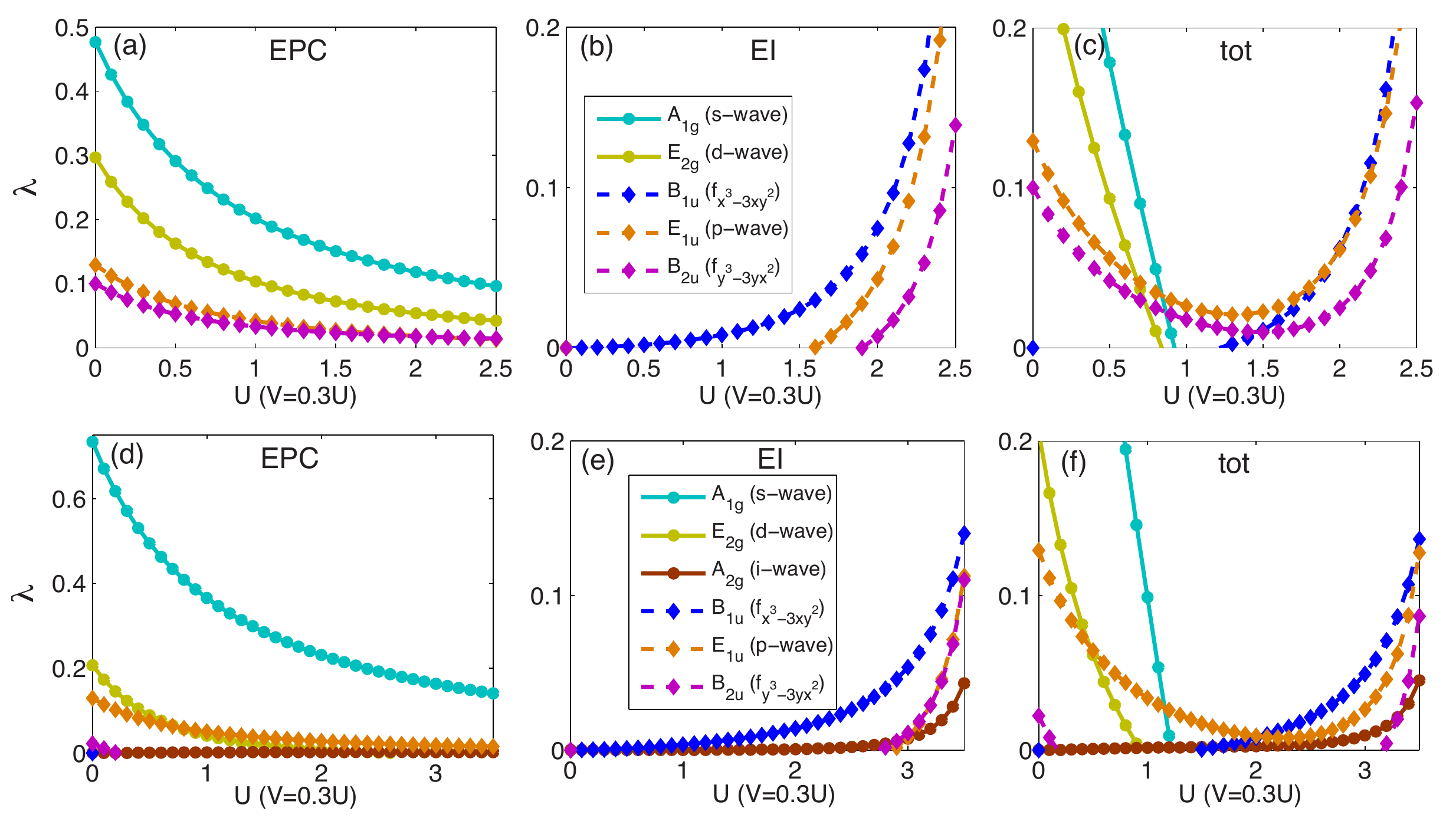}
\caption{This figure shows the pairing projection for a filling $\mu=0.08$ (a-c) and $\mu=-2.08$ (d-f) with only bubble contributions. The eigenvalues are scaled by a factor of $2/\sqrt{3}$.}
\label{fig_mu0p08mu2p08}
\end{center}
\end{figure}

The pairing eigenvalues from the projection method is shown in Fig.\ref{fig_mu0p08mu2p08} and the adopted lattice harmonics are provided in Table \ref{harmonics}. We find a good match between  Figs.\ref{fig_figure_bubbles} and \ref{fig_mu0p08mu2p08} in most regime. For both, $\mu=0.08$ and $\mu=-2.08$, the pairing eigenvalues from EPC in both method are very close for the leading and subleading channels. In addition, and importantly, both methods show that the non $s$-wave channles are more competitive respect to the isotropic $s$-wave symmetry for $\mu=0.08$ than for $\mu=-2.08$.

The results for $\lambda_i$'s from EI (see Fig.\ref{fig_figure_bubbles} (b) and Fig.~\ref{fig_mu0p08mu2p08}(b)) are also very similar, although the agreement is less quantitative than for the EPC case. For instance,
the subleading $\lambda_i$'s are more extended to lower $U$ in the eigenvalue calculation than in the projected one. This is because the eigenvalue calculation allows for a mixing of lattice harmonics with the same symmetry, which is not allowed in the projecting calculation. The most important difference is that $d$-wave symmetry is missing in the projection calculation, due again to the mixing harmonics in the eigenvalue calculation. Finally, the total case show also similar results (see Fig.\ref{fig_figure_bubbles}(c) and Fig.~\ref{fig_mu0p08mu2p08}(c)). Importantly, the dominant pairing states from the projection method are consistent with those from  solving the the linearized gap equation.

\begin{table*}[bt]
\caption{\label{harmonics}Adopted lattice harmonics for the gap function in the projection method.}
\begin{ruledtabular}
\begin{tabular}{cccc}
  irrep. & lattice harmonics ($F(\bm{k})$) & n-th NN pairing in kagome lattice &  \\
 \colrule
 $A_{1g}$(s-wave) &  $cosk_y+2cos(\sqrt{3}k_x/2)cos(k_y/2)$  & NN      \\
 $E_{2g}$(d-wave) &  $sin(\sqrt{3}k_x/2)sin(k_y/2)*\sqrt{3}$  &   NN    \\
  $E_{2g}$(d-wave) &  $cosk_y-cos(\sqrt{3}k_x/2)cos(k_y/2)*\sqrt{3}$  &   NN    \\
    $A_{2g}$(i-wave) &  $sin2k_y sin\sqrt{3}k_x-sin\frac{5}{2}k_y sin\frac{\sqrt{3}}{2}k_x-sin\frac{1}{2}k_y sin\frac{3\sqrt{3}}{2}k_x$  &      \\
   $E_{1u}$(p-wave) &  $sin(\sqrt{3}k_x/2)cos(k_y/2)*\sqrt{3}$  &   NN    \\
   $E_{1u}$(p-wave) &  $sink_y+cos(\sqrt{3}k_x/2)sin(k_y/2)$  &   NN    \\
   $B_{2u}$(f-wave) &  $sink_y-2cos(\sqrt{3}k_x/2)sin(k_y/2)$  &   NN    \\
   $B_{1u}$(f-wave) &  $sin(\sqrt{3}k_x)-2sin(\sqrt{3}/2k_x)cos(3k_y/2)$  &   NNN    \\
\end{tabular}
\end{ruledtabular}

\end{table*}

\begin{figure}[t]
\begin{center}
\includegraphics[width=0.48\textwidth]{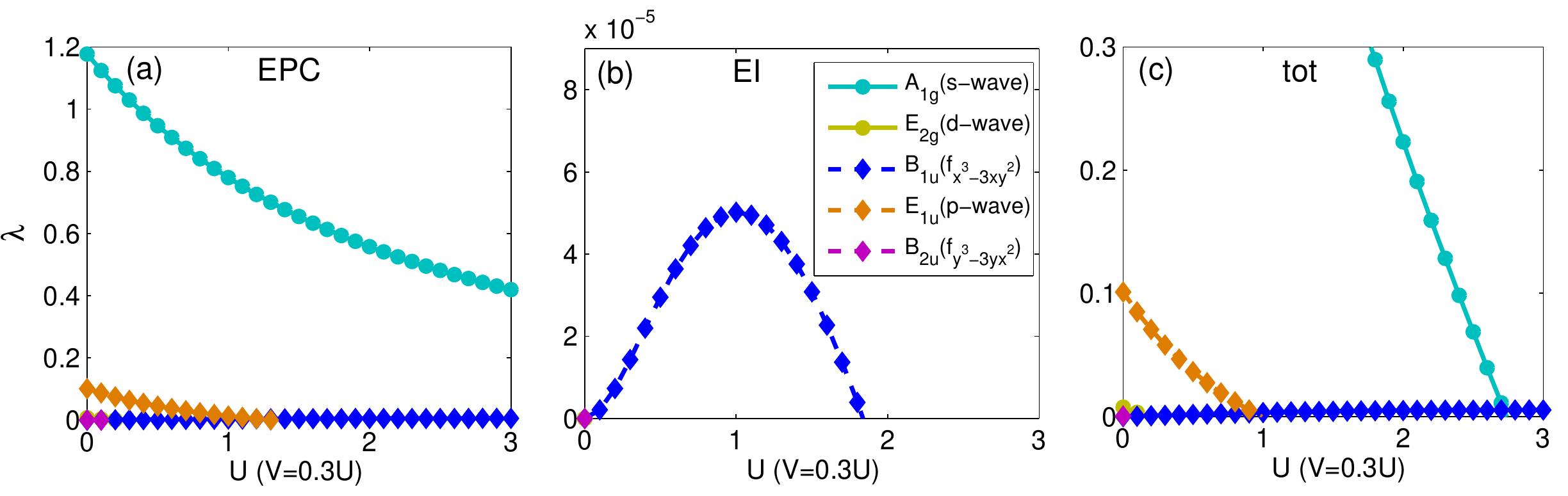}
\caption{
This figure shows the pairing projection for a filling $\mu=-3.5$ away from the van Hove fillings. The eigenvalues are scaled by a factor of $2/\sqrt{3}$.
}
\label{fig_mum2.5}
\end{center}
\end{figure}

Fig.\ref{fig_mum2.5} shows the case for $\mu=-3.5$, i.e., far from the two VHSs. The results are also similar to the results from the eigenvalues (Fig.\ref{fig_figureS1}). The $\lambda_i$'s for the EPC case are very similar qualitatively and quantitatively, and they dominate in the total case. In principle, the results for EI (Fig.\ref{fig_mum2.5} (b) and  Fig.\ref{fig_figureS1} (b)) seem to be different. While the projection calculation shows only the $B_{1u}$, the eigenvalue calculation shows more channels. However, the corresponding numbers are very small and four order of magnitude smaler than the $\lambda_i$'s for the EPC case.


\begin{figure}[t]
\begin{center}
\includegraphics[width=0.48\textwidth]{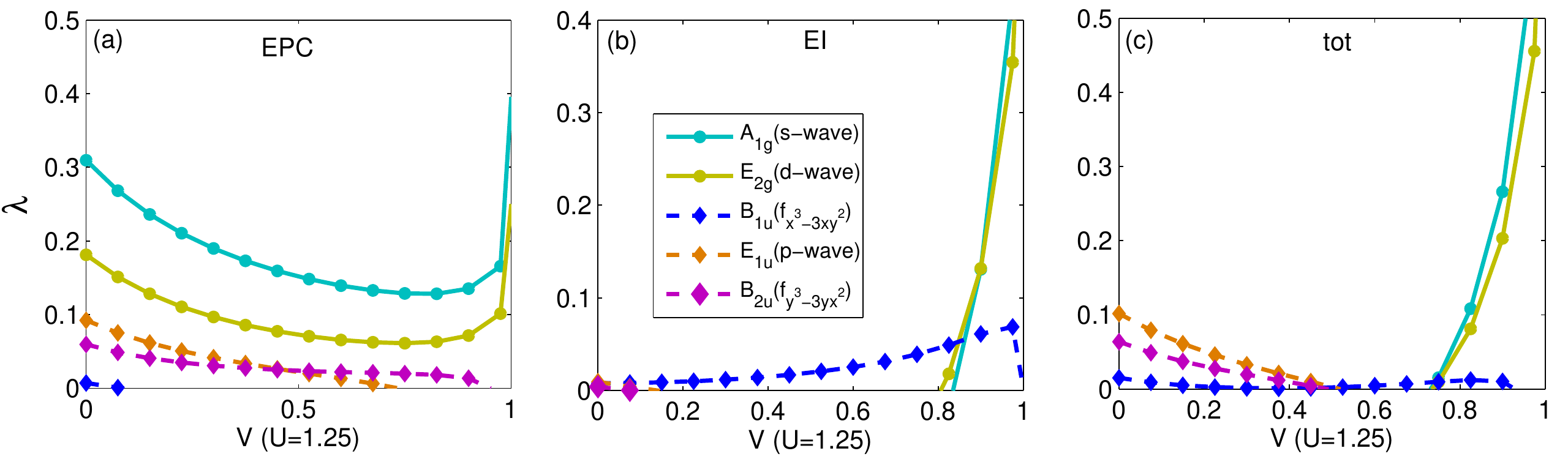}
\caption{
The same as Fig.\ref{fig_V} for the projection calculation. The eigenvalues are scaled by a factor of $2/\sqrt{3}$.
}
\label{fig_V_proj}
\end{center}
\end{figure}

Fig. \ref{fig_V_proj} shows results as a function of  $V$ calculated within the projection calculations which agree qualitatively with those presented in Fig. \ref{fig_V}.

We find the eigenvalues $\lambda_i$ from the linearized gap equation (Eq.~\eqref{eq:gapeigenvalues}) matches well in most regime with the $\lambda_i$ obtained from the projection method (Eq.\eqref{eq:lambdaiproj}), where nearest-neighbor (NN) and next NN harmonics are used in different pairing channels. While the largely used projection method\cite{Scalapino86} focus on one specific symmetry channel, the eigenvalues from the linearized gap equation may show a mixing between different symmetry channels. However, it is important to remark that the projection method fails to reproduce the behavior of the eigenvalues when ladders are included, which is interpreted as the eigenvectors in this case possess strong mixing with high harmonics.

\section{Gap functions and effective interactions from EPC and EI}

\begin{figure}[h!]
\begin{center}
\includegraphics[width=0.48\textwidth]{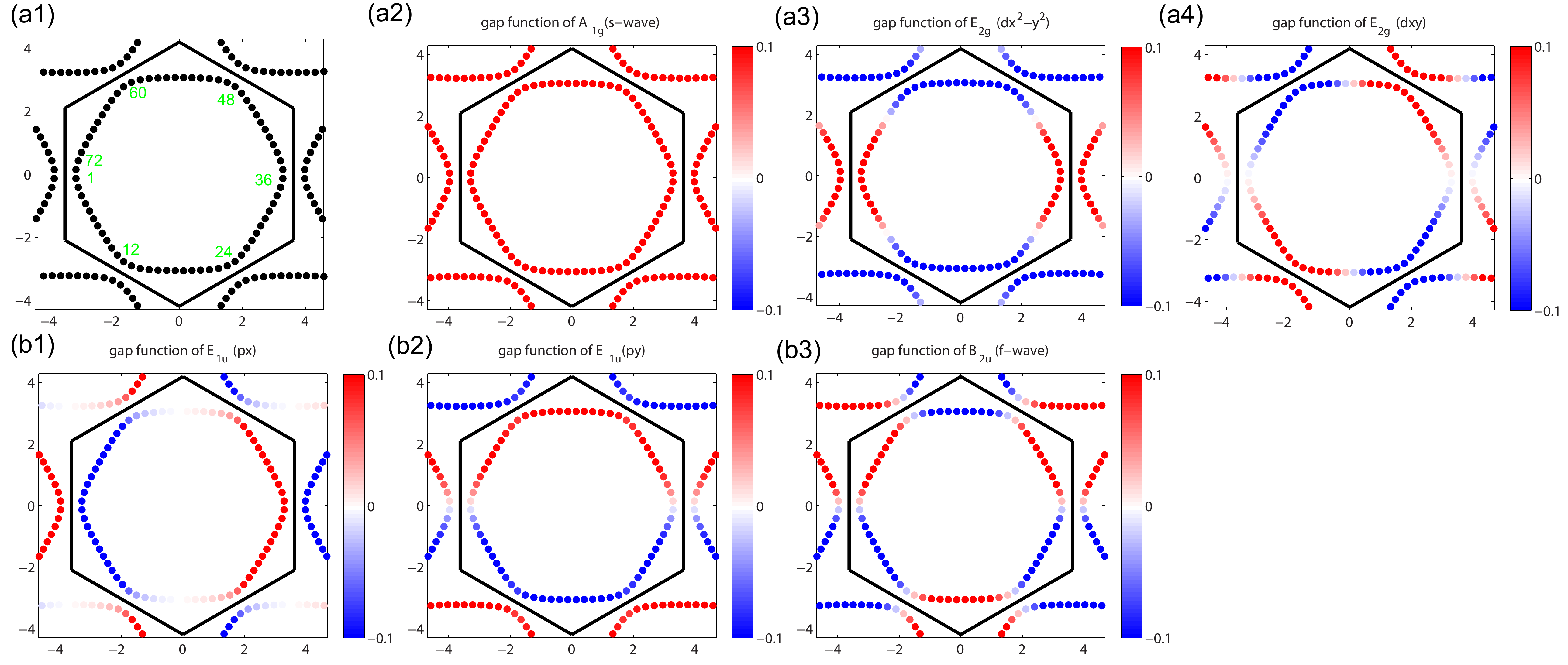}
\caption{
Fermi surface (a1) and gap functions for $A_{1g}$ and  $E_{2g}$ spin singlet pairing states (a2)-(a4), and  $E_{1u}$ and $B_{2u}$ spin triplet pairing states (b1)-(b3). The gap functions are from EPC without electronic interactions ($U=V=0$).
}
\label{fig_gapfun}
\end{center}
\end{figure}

\begin{figure}[h!]
\begin{center}
\includegraphics[width=0.48\textwidth]{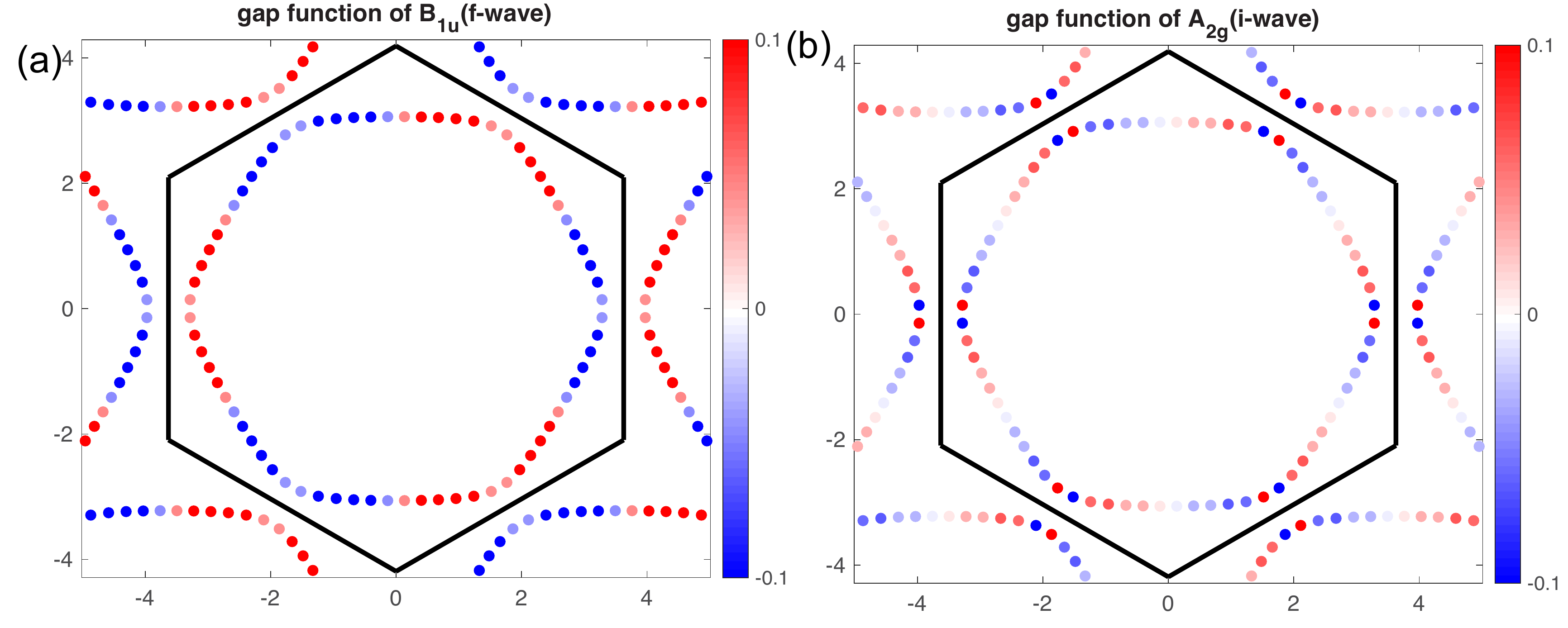}
\caption{
Gap functions for $B_{1u}$ pin triplet (a) and $A_{2g}$ spin singlet (b) pairing states. The $B_{1u}$ pairing is from the electronic interactions with $U=0.5$ and $V/U=0.3$ near the p-type VHS and the $A_{2g}$ pairing is from the electronic interactions with $U=1.5$ and $V/U=0.3$ near the m-type VHS.
}
\label{fig_gapfunVeff}
\end{center}
\end{figure}

\begin{figure*}[t]
\begin{center}
\includegraphics[width=0.8\textwidth]{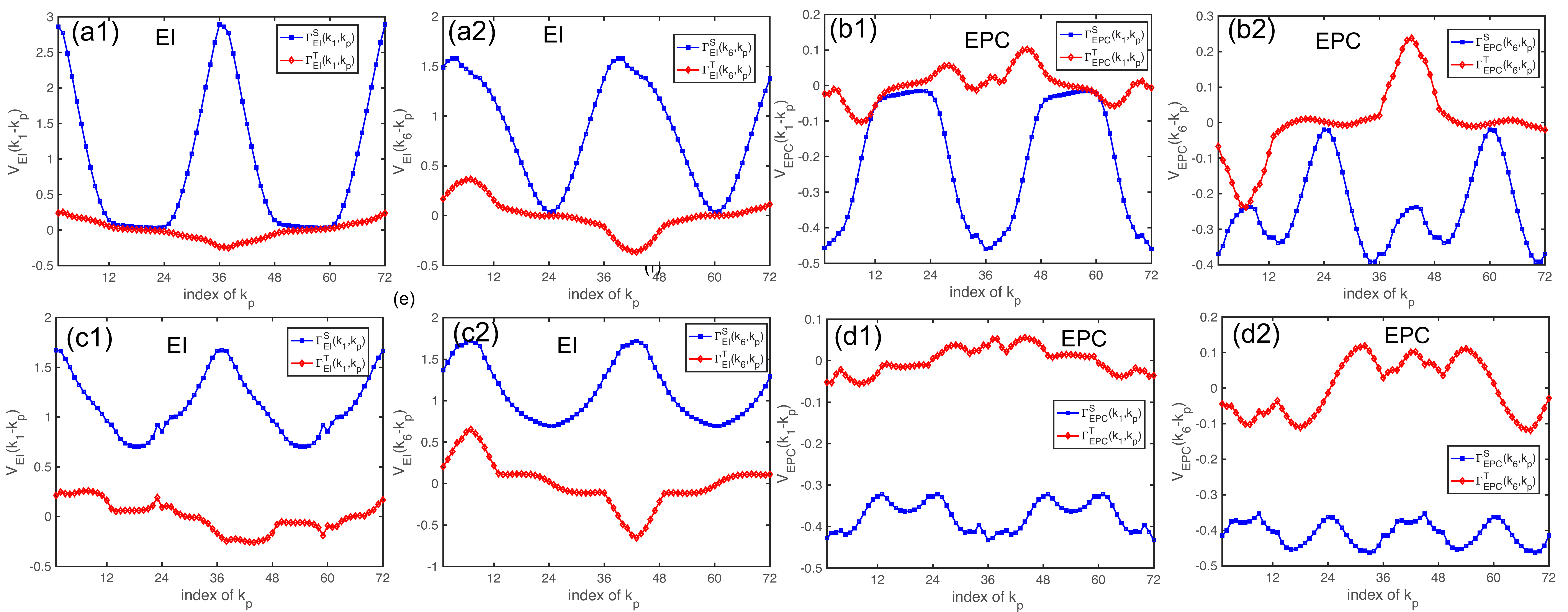}
\caption{
Effective pairing interaction $V(\bm{k}_{1/6},\bm{k}_p)$ in spin-singlet and spin-triplet channels from EPC and EI near the p-type VHS (top panels) and m-type VHS (bottom panels). The adopted interactions are $U=1.5$ and $V=0.3U$ in both cases.
}
\label{fig_pairST}
\end{center}
\end{figure*}

\begin{figure}[tb]
\begin{center}
\includegraphics[width=0.48\textwidth]{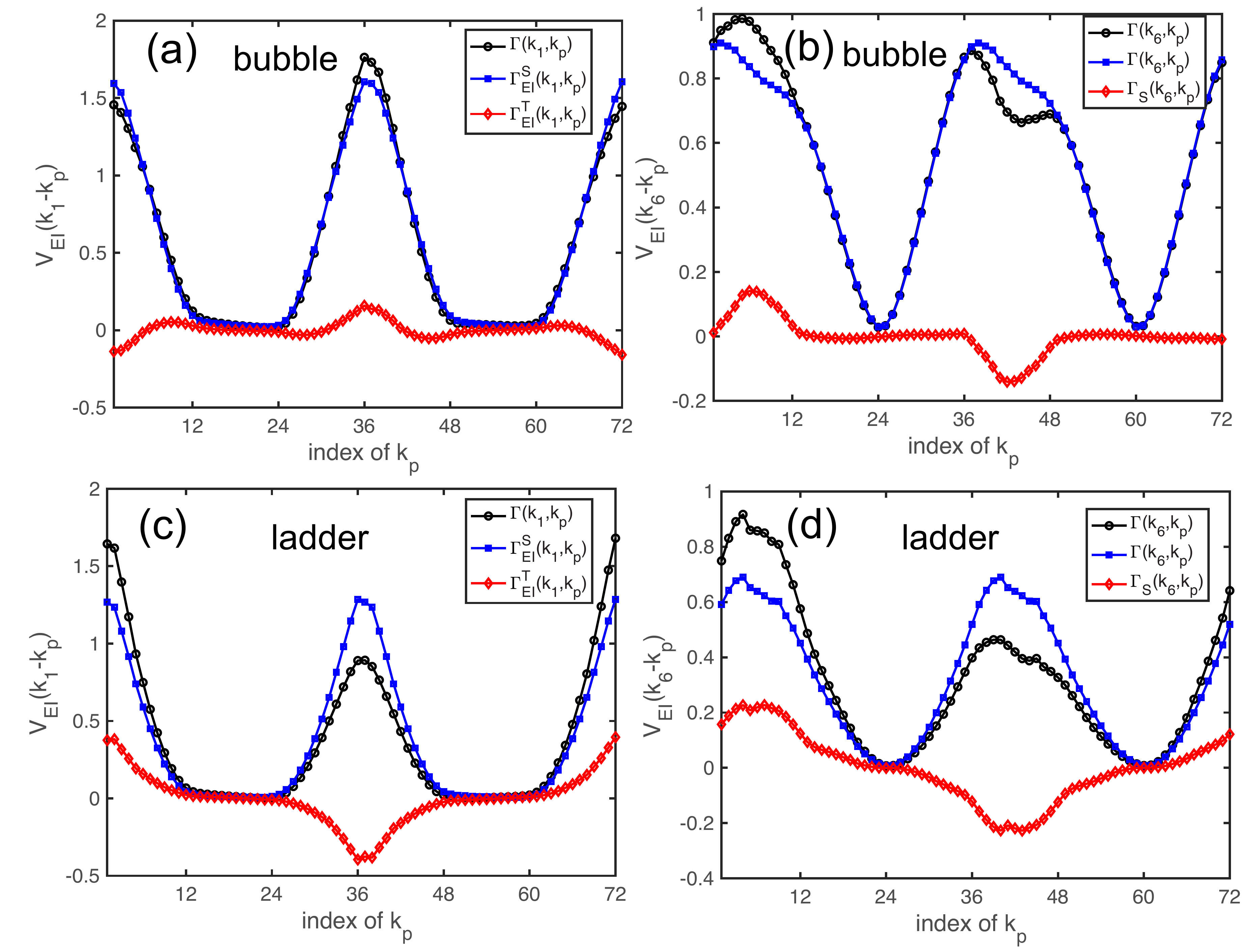}
\caption{
Effective pairing interaction $V(\bm{k}_{1/6},\bm{k}_p)$ in spin-singlet and spin-triplet channels from bubble and ladder diagrams near the p-type VHS. The adopted interactions are $U=1.5$ and $V=0.3U$.
}
\label{fig_pairBL}
\end{center}
\end{figure}

We display the representative gap functions in Fig.\ref{fig_gapfun} and Fig.\ref{fig_gapfunVeff}. Fig.\ref{fig_gapfun} shows the gap functions from the bare EPC without electronic interactions ($U=V=0$), while Fig.\ref{fig_gapfunVeff} displays the $B_{1u}$ and $A_{2g}$ gap functions from the electronic interactions. According to Ref.\cite{wu_PRL_21} and calculations from the projection method, the dominant pairing from EPC occur between the NN sites. We plot the effective interactions from EI and EPC in the Fig.\ref{fig_pairST} for the $p$-type and $m$-type VHS, respectively. For the case of $p$-type VHS, the interaction is anisotropic in both spin-singlet and spin-triplet channels, determined by the sublattice texture. The effective interaction in the spin-triplet channel $\Gamma^T_{EI}(\bm{k}_1,\bm{k}_{36})$ is attractive, slightly suppressing the spin-triplet pairing. The effective interaction from the EI bubbles and ladders are displayed in Fig.\ref{fig_pairBL} for $U=1.5$ and $V=0.3U$. We find that the interaction $\Gamma^{T}_{EI}(\bm{k}_1,\bm{k}_p)$ in the spin-triplet channel from the bubble and ladder diagrams are opposite. As the bubble contribution promotes the spin-triplet pairing (as discussed in Sec.5), the ladder contribution (dominated by the $U$ ladder terms) suppresses the spin-triplet pairing. When the ladder diagrams are included, the spin-triplet pairings are suppressed and spin-singlet pairings are promoted at large $U$ (as shown in Fig.3 in the main text).  For the $m$-type VHS, the anisotropy of the effective interactions is smaller, as shown in the bottom panels of Fig.\ref{fig_pairST}. Similarly, the ladder contribution also suppresses the spin-triplet pairing, generating comparable pairing eigenvalues in the spin-singlet and spin-triplet channels for large $U$.

\section{ Additional remarks  about present results and phenomenology of Kagome systems}

\subsection{The model, motivation, formalism and phonon softening }

Although several studies on the new kagome superconductors focus on correlation effects as the origin of the paring glue~\cite{wu_PRL_21,Tazai2022}, the interest on the role of phonons appears on the scene in recent time ~\cite{PhysRevLett.127.046401,PhysRevB.104.195130,2023arXiv230310080W,2022arXiv220702407Z,LiuHX2023} (to mention only a few papers). Remarkably, most of the electron-phonon studies on superconductivity focus on the usual $s$-wave paring, and for the estimation of $T_c$ they 
use the McMillan formula which requires the knowledge of the Coulomb pseudopotental parameter $\mu^*$.  In present paper we have investigated the role of the electron-phonon interaction in the context of the microscopic Hubbard-Holstein model, which is distinct from  the first-principle calculations approach ~\cite{PhysRevLett.127.046401,PhysRevB.104.195130,2023arXiv230310080W}.
The Holstein model is well known and considered as a  basic model for discussing superconductivity. It treats the vibration of the phonons in an average form, i.e., considering only one phonon frequency $\omega_D$ which can be associated with the Debye frequency. In addition, a bare and constant electron-phonon coupling $g_0$ is considered. In spite of their apparent simplicity, the model captures a realistic situation since the phonons and the electron-phonon interaction are rather constant at a bare level. In fact, the electron-phonon conventional superconductors are $s$-wave. In kagome lattice, we find a subdominant $d$-wave pairing with relatively large $\lambda$. With further inclusion of electronic interaction, $p$-wave and $f$-wave pairing can emerge in the crossover regime.

Since the electron-phonon coupling is in the moderate-coupling regime ($\lambda$ is moderate, see Ref.\cite{2022arXiv220702407Z}), we expect that our formalism can capture the leading pairing states and it may not change even when a more sophisticated method is adopted by solving
the full Eliashberg equations. The combined effect of EI and EPC in cuprates by solving the linearized gap equations has been earlier studied
and it was found that the EPC from the in-plane breathing motion of oxygen will suppress
the d-wave pairing~\cite{PhysRevB.54.14971}, which is qualitatively consistent with the
results from solving full Eliashberg equations~\cite{PhysRevB.54.12006}. This manifests that our results can be qualitatively consistent with those from solving the Eliashberg equations. T$_c$ can be accurately determined by solving the Eliashberg equations, which is beyond the scope of this work. 

Near van Hove fillings, phonons softening may become relevant but we expect the analysis of the pairing symmetries remains valid owing to the dominant sublattice interference. As the peaks in the susceptibility are mainly around the $\Gamma$ and M points (see Fig.2 in the main text), the phonon at the corresponding $\bf{q}$ points will get soft and this will accordingly enhance the EPC at these $\bf{q}$ points. However, as the EPC effective interaction is determined by the sublattice textures, this phonon softening will not alter the profile of EPC interactions (the EPC interaction for $k_1-k_p=Q_{2,3}$ is fixed to zero and phonon softening will not change this, as shown in Fig.2c).
Therefore, in our case, the phonon softening will only enhance the EPC interactions and the renormalized values of average electron-phonon coupling $\lambda$, which will not change the pairing symmetry and, at most, increases $T_c$.

\subsection{Realistic parameters and estimation of the superconducting critical temperature}

As discussed in the main text it is not our aim to reproduce the observed value of $T_c$. In the present stage of the topic the most important aspect is to identify pairing symmetry and its mechanism. Then, as in usual superconductivity, after that it is worthwhile to perform quantitative calculations for $T_c$. However, we can give a rough estimation for $T_c$. 

First, we recall that our chosen parameters are realistic. We choose $\omega_D=0.01t$ with  $t\sim 1$eV is of the order of $100-200$K~\cite{PhysRevLett.127.046401,PhysRevB.104.195130}.  By choosing $g_0 =0.1$, we obtain $\lambda \sim 0.5$ near the $p$-type van Hove in the $s$-wave channel (see Fig.~3(a) at $U=0$), which is in agreement with the more recent reported values~\cite{2023arXiv230310080W,2022arXiv220702407Z,LiuHX2023}. 
Note that the reported values of $\lambda$ in the first-principles calculations as well as in ARPES experiments is close to our $\lambda$ in the $s$-wave channel in absence of correlations, indicating that the adopted parameters are reasonable. 

At this point it is important to remark that in Ref.\cite{PhysRevLett.127.046401} electron-phonon superconductivity was ruled out because a low value for $\lambda \sim 0.3$ was obtained, which leads to a very low $T_c$ with $\mu^*=0.12$. However, as mentioned above, the recent literature shows that the electron-phonon coupling constant $\lambda$ is larger. 

Our calculations not only show that the value of $\lambda$ in the $s$-wave channel is close to the recent literature ($\lambda_s \sim 0.5$) but also show a large value of $\lambda$ in the $d$-wave channel ($\lambda_d \sim 0.3$) without correlations ($U=V=0$) due to the sublattice interference. This is in sharp contrast to other lattices. Since the value of $\lambda$ is  moderate, we can use the BCS formula $T^{s(d)}_c=1.14 \omega_D {\rm exp}(-1/(\lambda_{s(d)}-\mu^*_{s(d)}))$ to roughly estimate the transition temperature. Using our value for $\lambda_s$ and assuming $\mu^*_s \sim 0.1$ a value of $T_c^s$ of a few Kelvin can be expected. In addition, since $\mu^*_d$ is expected to be much lower than $\mu^*_s$~\cite{PhysRevB.73.060503}, $T_c^d$ of the order of a few Kelvin may also be expected.

Thus, due to the sublattice interference effects, in the kagome materials superconductivity with anomalous pairing may be expected even without or weak  electronic correlations which make the results rather robust and independent of the method of treating the electron-electron interaction. 

\subsection{Possible role of the CDW}

In the kagome superconductor AV$_3$Sb$_5$, superconductivity was originally observed inside the CDW order. It is well known that SDW and CDW can emerge in the kagome lattice. In our model the SDW or CDW instabilities occur at  high $U$ or $V$. So we are in a regime where SDW or CDW ordered states do not appear and do not influence the superconductivity. However, the competition between superconductivity and other ordered states is indeed an interesting future direction. In our paper, we focus on the superconductivity without a CDW order, as the CDW order in AV$_3$Sb$_5$ can be eliminated by external pressure and doping and is absent in the RERu$_3$Si$_2$ \cite{PhysRevMaterials.5.034803,ChinPhysLett.39.087401}, Ti-based kagome materials~\cite{YangHT2022}
 and Ta$_2$V$_{3.1}$Si$_{0.9}$~\cite{LiuHX2023}.

\end{document}